\documentclass[%
reprint,
 amsmath,amssymb,
 aps,
floatfix,
]{revtex4-2}

\usepackage{graphicx}
\usepackage{dcolumn}
\usepackage{bm}
\usepackage{hyperref}
\usepackage{booktabs}

\usepackage{geometry}
\usepackage{array}
\usepackage{colortbl}

\usepackage{lipsum} 
\usepackage{tabularx} 
\usepackage{multirow} 

\usepackage{quantikz}
\usepackage{tikz}
\usepackage{float}
\usepackage{enumitem}
\usepackage{makecell}

\begin{document}


\title{Hybrid Quantum-Classical Feature Extraction approach for Image Classification using Autoencoders and Quantum SVMs}

\author{Donovan Slabbert}
 \email{donovanslab@mweb.co.za}
\affiliation{Department of Physics, Stellenbosch University, Stellenbosch, 7600, South Africa \\
 }

\author{Francesco Petruccione}
 \email{petruccione@sun.ac.za}
\affiliation{National Institute of Theoretical and Computational Sciences (NITheCS), Stellenbosch, South Africa \\
 School of Data Science and Computational Thinking, Stellenbosch University, Stellenbosch, 7600, South Africa \\
 Department of Physics, Stellenbosch University, Stellenbosch, 7600, South Africa \\
}

\date{\today}

\begin{abstract}
In order to leverage quantum computers for machine learning tasks such as image classification, careful consideration is required: NISQ-era quantum computers have limitations, which include noise, scalability, read-in and read-out times, and gate operation times. Therefore, strategies should be devised to mitigate the impact that complex datasets, whether large-scale images or high-dimensional data, can have on the overall efficiency of a quantum machine learning pipeline. This may otherwise lead to excessive resource demands or increased noise. We apply a classical feature extraction method using a ResNet10-inspired convolutional autoencoder to both reduce the dimensionality of the dataset and extract abstract and meaningful features before feeding them into a quantum machine learning block. The quantum block of choice is a quantum-enhanced support vector machine (QSVM), as support vector machines typically do not require large sample sizes to identify patterns in data and have short-depth quantum circuits, which limits the impact of noise. The autoencoder is trained to extract meaningful features through image reconstruction, aiming to minimize the mean squared error across a training set of images. Three image datasets are used to illustrate the pipeline: HTRU-1, MNIST, and CIFAR-10. We also include a quantum-enhanced one-class support vector machine (QOCSVM) for the highly unbalanced HTRU-1 set, as well as classical machine learning results to serve as a benchmark. Finally, the HTRU-2 dataset is also included to serve as a benchmark for a dataset with well-correlated features. The autoencoder achieved near-perfect reconstruction and high classification accuracy for MNIST, while CIFAR-10 showed poorer performance due to image complexity, and HTRU-1 struggled because of dataset imbalance. This highlights the need for a balance between dimensionality reduction through classical feature extraction and prediction performance using quantum methods.
\end{abstract}

                              
\maketitle


\section{\label{sec:Introduction}Introduction\protect}

Quantum computing has led to the emergence of the field of quantum machine learning \cite{Biamonte2017, schuld2015introduction, schuld2021machine}, which blends together the concepts of classical machine learning and the principles of quantum mechanics \cite{nielsen2010quantum}. Quantum computing, and by extension quantum machine learning (QML), has challenges to overcome before its promised advantages or utility can be leveraged. Two of the most critical challenges facing quantum computing in the current NISQ-era are noise \cite{bharti2022noisy, cheng2023noisy} and runtime \cite{preskill2018quantum, chen2023complexity}. Quantum systems are susceptible to noise, which is accentuated at higher dimensions. Higher-dimensional circuits with many gates also tend to take much longer to run as compared to their classical counterparts. The need for scalable fault-tolerant quantum computers \cite{shor1996fault, preskill1998fault, divincenzo2000physical} and quantum error correction \cite{shor1995scheme, steane1996error, lidar2013quantum, knill1997theory} is only going to become more pressing, but in the meantime ways of reducing the dimension \cite{reddy2020analysis} of large datasets are crucial if they are to be encoded \cite{ghosh2021encoding} onto a quantum computer for machine learning purposes. This is especially true for image processing tasks, as encoding the entire image pixel-wise onto a quantum computer is not practical for large images.

A field where this is pertinent is astronomy. Observational datasets can be large and many of these datasets contain images \cite{wright2010wide, york2000sloan, dewdney2009square} . Astronomers already use machine learning for data preprocessing, analysis, predictions \cite{razim2021improving, banerji2010galaxy, fluke2020surveying, baron2019machine}, and more which means there will always be a need for more accurate and faster techniques or approaches. Quantum computing still has far to go before it can be a rival, let alone be a replacement, for classical machine learning. Despite this fact, QML can be applied to astrophysical and astronomical data on real devices or simulated to gain an indication of how well they can be expected to perform if the hardware were to become practical and fault tolerant in the future.

The chosen use case, similar to our earlier comparative study \cite{slabbert2024pulsar}, is that of pulsar data \cite{keith2010high, htru2_372, morello2014spinn}. A pulsars is a specific type of stellar remnant that forms after the death of a star at a sufficient solar mass \cite{lyne2012pulsar}, but not enough mass to become a black hole. Pulsars are characterized as highly dense objects where the total collapse into a black hole is avoided only because the object's self-gravity cannot overcome neutron degeneracy pressure \cite{kippenhahn2012stellar}. Identifying pulsars is important and forms part of astronomical surveys, as well as investigations with respect to gravitational wave detection and pulsar astronomy \cite{hulse1975discovery, stairs2003testing, foster1990constructing, mclaughlin2013north, verbiest2016international}.

This paper focuses on image classification of the three-channel HTRU-1 \cite{keith2010high, morello2014spinn} image dataset through a quatum-enhanced support vector machine (QSVM) \cite{MariaSchuld2021, schuld2023supervised}. We include classical machine learning results and a quantum-enhanced one-class support vector machine {QOCSVM} for anomaly detection \cite{chandola2009anomaly}. This was done since the HTRU-1 set is highly imbalanced. The minority class (pulsars) only make up $2\%$ of the dataset. We also include this for the MNIST \cite{lecun1998gradient} and CIFAR-10 \cite{krizhevsky2009learning} datasets and apply the chosen methods to the HTRU-2 dataset \cite{lyon2016fifty, htru2_372} as well.

In order to encode images, specifically three-channel images from HTRU-1 and CIFAR-10, onto a quantum device, amplitude encoding \cite{pande2024comprehensive, schuld2021effect} is jointly implemented with a classical feature extraction and dimensionality reduction pipeline. Feature extraction methods include edge detection \cite{ziou1998edge}, histogram of oriented gradients (HOG) \cite{dalal2005histograms}, principal component analysis (PCA) \cite{wold1987principal, pearson1901liii}, convolutional neural networks (CNNs) \cite{lecun1998gradient, jogin2018feature} etc. The chosen feature extraction scheme is a classical ResNet10-inspired convolutional autoencoder (CAE), trained for image reconstruction \cite{hinton2006reducing}. The latent space is isolated and flattened after training for use in our machine learning tasks. QSVMs and QOCSVMs typically require only a few samples to learn the pattern in the dataset \cite{schuld2023supervised, slabbert2024pulsar}, however, the trade-off is quadratic scaling. In order to fit a SVM to a dataset, each sample in the training set has to be compared to every other sample. This is relevant both during training and testing.

The rest of the paper is structured as follows: Section 2 will briefly explain anomaly detection and the chosen QML methods. It will also explain the classical feature extraction pipeline. Section 3 will discuss the methods in more detail. Data preprocessing steps, choice of hyperparameters etc. can be found here. Section 4 is the results section and includes both the anomaly detection and supervised classification results. Section 5 is the conclusion. Two appendices are included: Appendix A gives more information on the metrics used for performance evaluation and Appendix B outlines the model architectures.

\section{\label{sec:Theory}Theory}

\subsection{\label{sec:Anomaly detection}Classification and anomaly detection}

Classification is a machine learning task where the goal is to accurately predict the class labels of two or more classes in a dataset. Anomaly detection \cite{chandola2009anomaly} comes in different forms, such as time-series anomaly detection \cite{blazquez2021review}, outlier detection \cite{hodge2004survey}, and minority sample prediction \cite{he2009learning}. Some approaches to anomaly detection include clustering-based methods \cite{lima2010anomaly}, density estimation methods \cite{nachman2020anomaly}, statistical methods \cite{aissa2016semi}, isolation forests \cite{liu2008isolation} etc. In our case, supervised anomaly detection, it is similar to classification in the sense that the goal is to accurately predict or flag a sample in a test set as an anomaly (belonging to the minority class). There are key differences between standard classification and anomaly detection, including the imbalance in the dataset and how the model is trained. The class imbalance is heavily skewed towards the samples that are considered normal. Training the model involves only using normal data. Anomalies are only introduced during testing. The model then identifies anomalous samples based on deviations beyond a certain threshold from normal behavior. Since we have ground truth in the form of labeled data, the anomaly detection we implement can be classified as supervised anomaly detection. From now on we may call samples belonging to the minority class anomalies.

Consider the following equation describing supervised classification and anomaly detection \cite{schuld2021machine}:

\begin{equation}
f_{\theta}(\vec{x}) = y \quad | \quad f_{\theta}:\bm{X} \rightarrow \bm{Y}, \quad \vec{x} \in \bm{X}, \quad y \in \bm{Y}.
\end{equation}

Here, $\bm{X}$ and $\bm{Y}$ represent the input and output domains, respectively. $\vec{x}$ denotes an $n$-dimensional feature vector encompassing $n$ features. The label $y$ belongs to the set ${-1, 1}$, where $1$ designates normal samples and $-1$ denotes anomalies. The function $f_{\theta}$ signifies the optimized or trained model, with $\theta$ representing any parameters that need to be optimized during training.

To evaluate the performance of a supervised anomaly detection or classification approach, we use a test set, which must include both normal and anomalous samples. The effectiveness of the anomaly detection model can be assessed using many metrics, where we choose to use the following ten metrics \cite{powers2020evaluation, tharwat2020classification, chicco2020advantages}: accuracy, proportion of postive predictions, precision, recall, negative prediction value, specificty, false positive rate, false negative rate, F1 score and Matthew's Correlation Coefficient. We include all ten for a holistic comparison, but the most important metrics focus on the positive class, which in this use case are the anomalies (pulsars). Precision and recall do just this: precision measures the proportion of correctly predicted anomalies from all samples predicted as anomalies and recall measures the proportion of correctly predicted anomalies from all anomalous samples in the test set. Proportion of positive predictions (PPP) is just the amount of anomalies predicted, correct or otherwise. This value has to be around $2\%$ for the HTRU-1 set, but will change depending on the dataset. The other metrics are discussed in Appendix \ref{sec:performance_metrics}.

\begin{table}[t]
\centering
\renewcommand{\arraystretch}{1.4}
\begin{tabular}{p{0.3\linewidth} p{0.6\linewidth}}
\toprule
\textbf{Metric} & \textbf{Equation} \\
\midrule
Accuracy & $\frac{\text{TP} + \text{TN}}{\text{TP} + \text{FP} + \text{TN} + \text{FN}}$ \\
PPP & $\frac{\text{TP} + \text{FP}}{\text{TP} + \text{FP} + \text{TN} + \text{FN}}$ \\
Precision & $\frac{\text{TP}}{\text{TP} + \text{FP}}$ \\
Recall & $\frac{\text{TP}}{\text{TP} + \text{FN}}$ \\
NPV & $\frac{\text{TN}}{\text{TN} + \text{FN}}$ \\
Specificity & $\frac{\text{TN}}{\text{TN} + \text{FP}}$ \\
FPR & $\frac{\text{FP}}{\text{FP} + \text{TN}}$ \\
FNR & $\frac{\text{FN}}{\text{FN} + \text{TP}}$ \\
F1 Score & $2 \times \frac{\text{Precision} \times \text{Recall}}{\text{Precision} + \text{Recall}}$ \\
MCC & $\frac{\text{TP} \times \text{TN} - \text{FP} \times \text{FN}}{\sqrt{(\text{TP} + \text{FP})(\text{TP} + \text{FN})(\text{TN} + \text{FP})(\text{TN} + \text{FN})}}$ \\
\bottomrule
\end{tabular}
\caption{Evaluation metrics expressions for classiciation and supervised anomaly detection. TP, TN, FP, and FN represent the true positives, true negatives, false positives, and false negatives, respectively, after applying a supervised anomaly detection model to a test set.}
\label{table=equations}
\end{table}

\begin{figure*}[t]
  \centering
  \includegraphics[width=\textwidth]{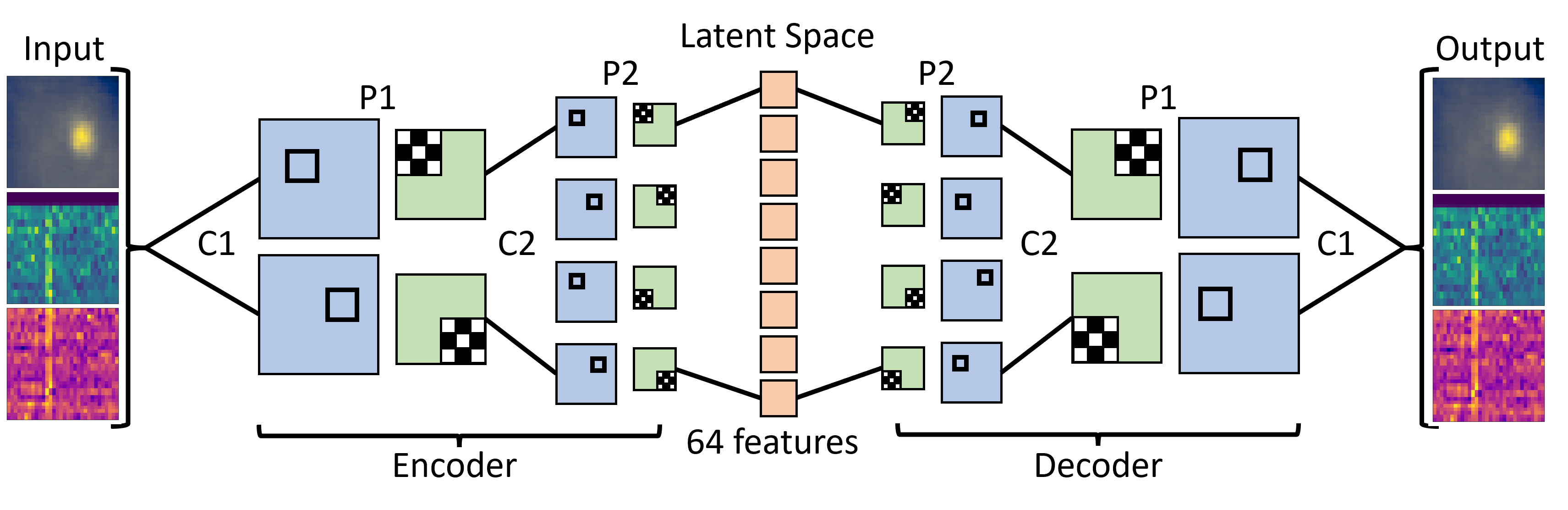}
  \caption{Image reconstruction pipeline made possible through a convolutional autoencoder (CAE). The goal of the autoencoder is to learn how to reconstruct images and ultimately to serve as an abstract feature extraction scheme for use in later quantum machine learning blocks. This figure does not represent the true architecture of the model. P and C indicate pooling and convolution layers.}
  \label{fig:figure1}
\end{figure*}

\subsection{\label{sec:SVM} Support Vector Machines}

Standard support vector machines (SVMs) \cite{noble2006support, boswell2002introduction} are best described as maximal margin classifiers that work by mapping input features to a higher-dimensional feature space through a feature map. In this space, a separating hyperplane is determined using important samples, known as support vectors, which help define the orientation and location of the hyperplane. 

The decision boundary in the feature space can be expressed as:

\begin{equation}
\vec{w} \cdot \phi(\vec{x}) - \rho = 0
\end{equation}

where $\vec{w}$ is the weight vector, $\phi(\vec{x})$ is the feature map, and $\rho$ is the offset (bias). The hyperplane is defined by the set of points $\vec{x}$ where this equation holds.

To compute the similarity between data points without explicitly mapping them to the higher-dimensional space, SVMs utilize kernel functions. This method, known as the kernel trick \cite{scholkopf1999advances, boser1992training}, allows the SVM to efficiently operate in the higher-dimensional space. The choice of kernel influences the position and orientation of the hyperplane, enabling the SVM to separate data even when the relationship between features and class labels is complex and non-linear.

\subsubsection{\label{sec:OCSVM} One-Class Support Vector Machines}

One-class support vector machines (OCSVMs) \cite{scholkopf2002learning, scholkopf2001estimating, muller2018introduction, hejazi2013one} extend the principles of standard SVMs by training exclusively on normal samples. During testing, anomalous samples (pulsars) are introduced, and anomalies are flagged based on their position in feature space relative to a separating boundary. 

In feature space, all normal data typically fall within this boundary. Samples outside of it are considered anomalous. The OCSVM utilizes a decision function to assess the distance from each point to the separating hyperplane, with any value greater than \(0\) indicating a normal sample and any value less than \(0\) indicating an anomaly. This distance is called the anomaly score.

The OCSVM is trained solely on normal samples, represented by the training set \(\mathcal{D}_{\text{train}} = \{\vec{x}_i \mid y_i = 1\}\), to learn the boundary that separates normal behavior from anomalies. The decision function \(f_{\theta}(\vec{x})\) maps the input space \(\mathbb{R}^n\) to a signed distance in \(\mathbb{R}\), defining a hyperplane in feature space. During testing, this function is given by

\begin{equation}
f_{\theta}(\vec{x}) = \vec{w} \cdot \phi(\vec{x}) - \rho,
\end{equation}

with a sample \(\vec{x}\) classified as normal if \(f_{\theta}(\vec{x}) \geq 0\) and as an anomaly if \(f_{\theta}(\vec{x}) < 0\).

In contrast, standard SVMs are trained on labeled data from two classes, aiming to find the optimal hyperplane that separates these classes with maximum margin. The decision function in standard SVMs closely resembles that of OCSVMs, focusing on which side of the hyperplane a sample falls on, rather than whether it lies inside or outside of it. This decision function is referred to as the decision score. Unlike OCSVM, standard SVMs utilize both positive and negative samples during training.

\subsubsection{\label{sec:QK} Quantum Kernel}

Quantum kernels refer to kernel functions estimated or determined on a quantum computer \cite{huang2021power, liu2021rigorous} where the feature space corresponds to the quantum mechanical Hilbert space after feature encoding \cite{schuld2019quantum, havlivcek2019supervised}. To implement a quantum-enhanced SVM and OCSVM, we utilize a quantum kernel. We try to keep the quantum circuit as simple as possible, to limit depth-related noise. This is achieved by minimizing gate-based noise that would be introduced if the circuit were run on real devices, thereby also limiting computational costs. We extend a known method \cite{MariaSchuld2021, schuld2021supervised} by amplitude encoding one feature vector and then applying a second encoding block, which is simply the complex conjugate transpose of the first encoding block applied to a second feature vector. We express amplitude encoding as \cite{ghosh2021encoding, pande2024comprehensive}:

\begin{equation}
    |\phi\rangle = S(\vec{x})|00\ldots0\rangle = \sum_{i=0}^{2^n-1} x_i |i\rangle,
\end{equation}

where the features from the feature vectors are embedded into the amplitudes of a $2^{n}$ dimensional quantum state. Amplitude encoding is only possible if

\begin{equation}
    \sum_{i=0}^{2^n-1} |x_i|^2 = 1,
\end{equation}

which implies that all feature vectors should be normalized before amplitude encoding. This ensures that the quantum state represents a valid physical state according to the requirements of quantum mechanics. \cite{nielsen2010quantum}

Next, we measure with respect to a projector matrix. The projector matrix approach allows us to compute the quantum kernel, or the inner product between quantum states, directly. This eliminates the need for a SWAP test \cite{MariaSchuld2021, schuld2021supervised}, which is a common but resource-intensive method for estimating state overlap that utilizes ancilla qubits \cite{nielsen2010quantum}. By measuring with respect to the projector matrix, we simplify the computation and efficiently evaluate kernels or similarities between feature vectors encoded in the quantum states.

The kernel function can be expressed as \cite{MariaSchuld2021}:

\begin{equation}
    K(\vec{x},\vec{x}')|\bra{00\ldots0} S^{\dagger}(\vec{x}')S(\vec{x})\ket{00\ldots0}|^2,
\end{equation}

where both encoding operators have been applied and the measurement is performed with respect to the projector matrix

\begin{equation}
    \rho = |00\ldots0\rangle \langle 00\ldots0|,
\end{equation}

which is also the zero state density matrix.

\subsection{\label{sec:Autoencoder} Classical Autoencoder}

\begin{figure*}[t]
  \centering
  \includegraphics[width=\textwidth]{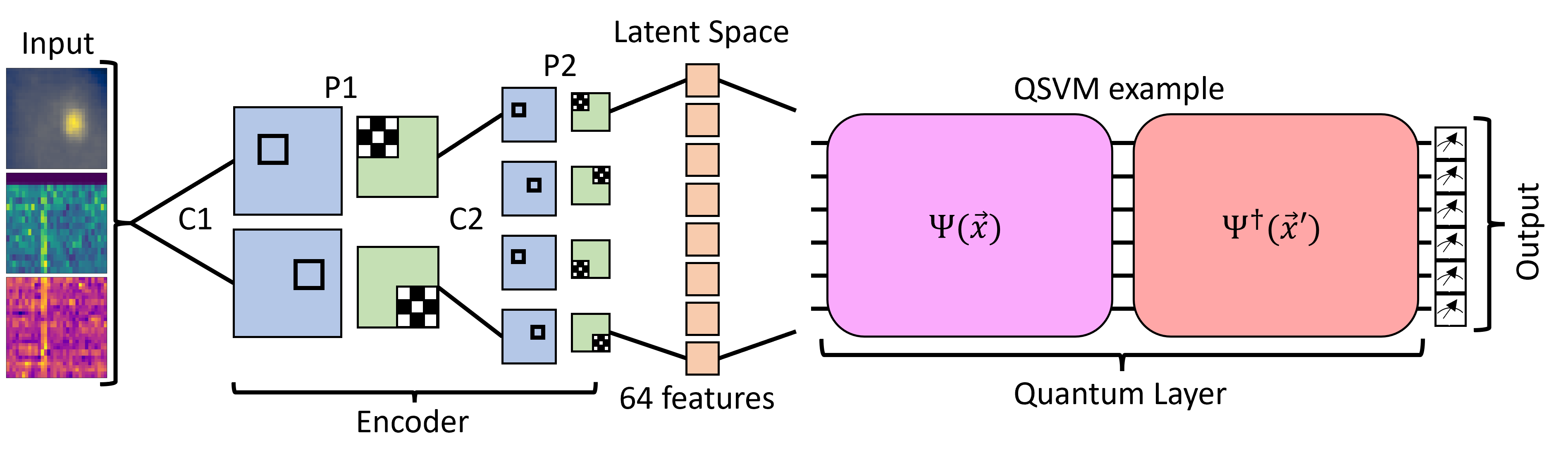}
  \caption{Concept Figure: The isolated convolutional encoder segment of the classical autoencoder used to encode the input images into the latent space. P1, P2, C1, and C2 indicate the pooling and convolutional steps, but do not represent the true architecture of the model. The flattened latent space leads to 64 abstract features that serve as input into the quantum block through amplitude encoding.}
  \label{fig:figure2}
\end{figure*}

Classical autoencoders (CAEs) \cite{hinton2006reducing} are similar to standard neural networks or convolutional neural networks (CNNs). In these networks \cite{goodfellow2016deep}, nodes are interconnected via weights, which determine the strength of the connections between neurons. These weights are learned through an iterative optimization procedure, typically using a gradient descent algorithm via forward and back propagation. The learning process aims to minimize a loss function that measures the difference between the input and the reconstructed output, ensuring that the network accurately captures the essential features of the input data.

The primary goal of autoencoders is to abstract the input data into a set of feature maps: compact representations of the input that capture its most relevant characteristics and spatial information. This abstraction process occurs during the encoder phase of the autoencoder, where the input data is progressively transformed into a lower-dimensional representation known as the latent space. The latent space serves as a compressed version of the input data. It contains the most critical information needed to reconstruct the original input.

Once the data is encoded into the latent space, the abstract feature maps are passed through the decoder, the remaining component of the autoencoder network. The decoder's goal is to take the abstract representations in the latent space and decode them back into the original input format. The effectiveness of an autoencoder is measured by how well the decoded output matches the initial input data, with the best autoencoders producing outputs nearly indistinguishable from the original inputs. The idea can be visualized as seen in Figure \ref{fig:figure1}.

CAEs operate through convolution operations, which are the core building blocks of convolutional neural networks. These operations involve matrix multiplication, where a learnable filter or kernel scans over the input image. The filter is applied to small sections of the image at a time, capturing local patterns such as edges, textures, and other features. The result of these operations is a set of feature maps or matrices that represent the presence of specific features in different regions of the image. These feature maps are then passed through successive layers, leading to increasingly abstract and complex representations of the original image features.

Beyond image reconstruction, autoencoders are also valuable for feature extraction, where the encoder is used to learn compact representations of input data. These representations can be used as input features for other machine learning models, enabling more efficient and accurate data analysis. This supports why we decided to utilize autoencoders for feature extraction. Convolutional Autoencoders (CAEs) are typically designed with convolutional layers, which capture spatial hierarchies in the data, and pooling layers, which reduce the dimensionality of the feature maps.

\section{\label{sec:Methodology} Methodology}

\subsection{\label{sec:datasets}Datasets and feature extraction}

As mentioned in the introduction, we will be considering three datasets as part of the feature extraction and anomaly detection pipeline. These are the HTRU-1 pulsar, MNIST, and CIFAR-10 image datasets. Additionally, the tabular HTRU-2 dataset is included for benchmarking. All datasets are available in the data availability statement.

\subsubsection{HTRU-1}

The HTRU-1 image dataset contains 60,000 samples, each of which is a three-channel image of 32x32 pixels. The three channels are similar to RGB values, but instead represent other quantities for each sample: Channel 1: Period Correction - Dispersion Measure Surface; Channel 2: Phase - Sub-band Surface; Channel 3: Phase - Sub-integration Surface. 

The dataset is divided into a training set of 40,000 samples and test and validation sets of 10,000 samples each, from which subsets will be sampled for training and testing purposes in the quantum blocks. Only $2\%$ of this dataset represents anomalies (pulsars). The CAE for image reconstruction is trained using the entire training set. Data preprocessing includes min-max-scaling and data augmentation. The data augmentation implemented are random horizontal and vertical flips and random rotations up to a maximum of 15 degrees.

\subsubsection{HTRU-2}

The HTRU-2 dataset consists of 17,898 samples of eight features each. Approximately $9\%$ of the dataset consists of anomalies (pulsars). The results from this dataset will serve as a benchmark to compare the results on the image dataset above, since it is assumed that the features from HTRU-2 are all meaningful. We do not have to apply the feature extraction scheme here, as the samples are not images.

\subsubsection{MNIST}

MNIST is a gray scale image dataset of 28x28 images of the numerical digits 0-9. We isolate the first two digits, 0 and 1, creating a separate dataset for binary classification to serve as a benchmark for both the image reconstruction scheme and comparison with quantum methods.

\subsubsection{CIFAR-10}

CIFAR-10 is similar to HTRU-1, being three-channel 32x32 pixel images. There are 50,000 samples in the training set and 10,000 in the test set, which contains 10 classes of everyday objects. Similar to MNIST, we isolate the first two classes (airplane and automobile) for binary classification, serving as a second benchmark for image reconstruction.

\subsubsection{CAE Feature Extraction}

The chosen CAE architecture, designed for feature extraction and trained via image reconstruction, should balance model complexity and generalization. More complex and deeper models are often capable of reconstructing images with higher fidelity, but they could also be more prone to overfitting \cite{zhang2021understanding}. This is when the model learns noise or irrelevant details and essentially memorises the training set. Vanishing gradients, where the learning process stagnates and fails to make meaningful progress \cite{hochreiter1997long}, is also a consideration. It is important to note that sometimes the model has to be adapted depending on the input that it will receive.

Included in these considerations are the points that were mentioned previously: we will be isolating the latent space for input into a quantum block. Since quantum computing has slow data read-in times and SVMs in general scale quadratically, it is imperative to keep the latent space and total sample amount at a sufficiently low, yet effective number to complete the task. In Figure \ref{fig:figure2} the idea of isolating the latent space and replacing the decoder with the quantum block is illustrated. This approach is similar to using a hybrid convolutional neural network and variational quantum circuit in conjunction for similar purposes \cite{wang2023quantum, senokosov2024quantum}.

All autoencoders were designed in such a way that the latent space, when flattened, contained precisely 64 abstract features for efficient encoding using amplitude embedding. 64 features require a 6-qubit system in order for amplitude embedding to accommodate all features. More features in the latent space would imply both an improved image reconstruction and feature extraction performance, which would ultimately lead to an improved performance in the quantum block results. We chose 64 features, as stated before, specifically to find a balance between computational complexity and performance. This is due to the fact that quantum machine learning circuits, especially for quadratically scaling circuits for QSVMs (and QOCSVMs), can have large runtimes. Our goal was to assess whether a low-dimensional latent space with a small number of features would suffice for a robust model.

\subsection{\label{sec:Specifics} Specifics}

\subsubsection{\label{sec:Details} Autoencoder, QSVM and QOCSVM details}

We initially started with a really simple CAE architecture of three convolutional layers in the encoder and decoder each. In the encoder, the images would progressively be scaled down in dimensionality to a latent space of precisely 64 features. The decoder would then be the inverse of the encoder, trained with its own set of parameters to decode the latent space back into the initial image. This architecture was too simple for the HTRU-1 dataset, which led to the exploration of pre-trained ResNet18 models \cite{he2016deep} for the encoder part and custom decoders designed to fit into the latent space of ResNet. ResNet18, although extensive, was overly complex. We settled for the middle ground of a ResNet10-inspired architecture \cite{gong2022resnet10}, which in essence means that the architecture of ResNet10 was taken as the base skeleton architecture for the encoder, but changes were made to it depending on the need. Changes would include kernel sizes, stride, number of input/output channels per layer, among other changes.

The final architecture layer layout chosen is illustrated in Table \ref{table:encoder_decoder} in Appendix \ref{sec:Autoencoder Architecture}. The loss function chosen was mean squared error loss. We considered using the structural similarity index measure (SSIM) \cite{wang2004image}, but the HTRU-1 images are too abstract for a more visual measure. The optimizer for forward and back propagation during the image reconstruction training was set to the Adam optimizer at a learning rate of $0.001$. We also implemented built in L2 regularization using a weight decay of $1\times 10^{-5}$. The training set was split further into a ratio of 80:20 for a new training and validation set, for validation purposes. The convergence of the validation loss would serve as an indication of overfitting \cite{goodfellow2016deep}. Finally, training was done in batches of 256 samples per epoch and we set the number of epochs at $20$, as this was the point of loss convergence for HTRU-1. After the image reconstruction models were trained, we extracted features from the entire dataset by sending it through the encoder. At this point sub-sampling was introduced for a smaller, but representative set through stratification to the original label distribution, for feeding into the quantum block. There was separation made for anomaly detection and classification training sets here, since anomaly detection may only be trained on normal samples.

We used scikit-learn as the Python library for fitting the quantum kernel SVMs to the training data before applying it to a test set. In all SVM runs, this includes both classical and quantum results. We introduced class weight parameters of  $\{-1:50, 1:1.02\}$ for HTRU-1 runs, to weigh the anomalous samples more during training. Recall that a label of 1 represents normal samples and -1 anomalous samples. The class weights are set specifically to reflect the 2:98 ratio of anomalous samples to normal samples in the HTRU-1 set and are calculated by the inverse frequency of class appearances. This was changed to $\{-1: 11.11, 1: 1.10\}$ for HTRU-2 runs and was removed entirely for MNIST and CIFAR-10 runs due to their balanced class distributions. The class ratios are 9:91 for HTRU-2 and practically 50:50 for MNIST and CIFAR-10. We did similarly for the OCSVM runs by introducing a 'nu' parameter of 0.2 for HTRU-1 and 0.9 for HTRU-2. We did not apply OCSVMs to MNIST or CIFAR-10, since these sets are essentially balanced. The 'nu' parameter controls the proportion of data considered outliers and the minimum fraction of support vectors, balancing the model's strictness in detecting anomalies. The choice of kernel for the classical runs was the standard radial basis function.

Finally we chose a sub-sample of 500 training and testing samples for training and testing the quantum blocks. This is different from the training and test sets during the image reconstruction. The shape of the input into the SVMs is therefore (500 samples, 64 features) which lead to a 6-qubit quantum state when amplitude encoded for quantum runs. We chose 500, since it seemed to be the trade-off point between computational demand and number of samples for pattern learning.

\subsubsection{\label{sec:Channel-based approach}Channel-based approach for HTRU-1}

\begin{figure*}[t]
  \centering
  \includegraphics[width=\textwidth]{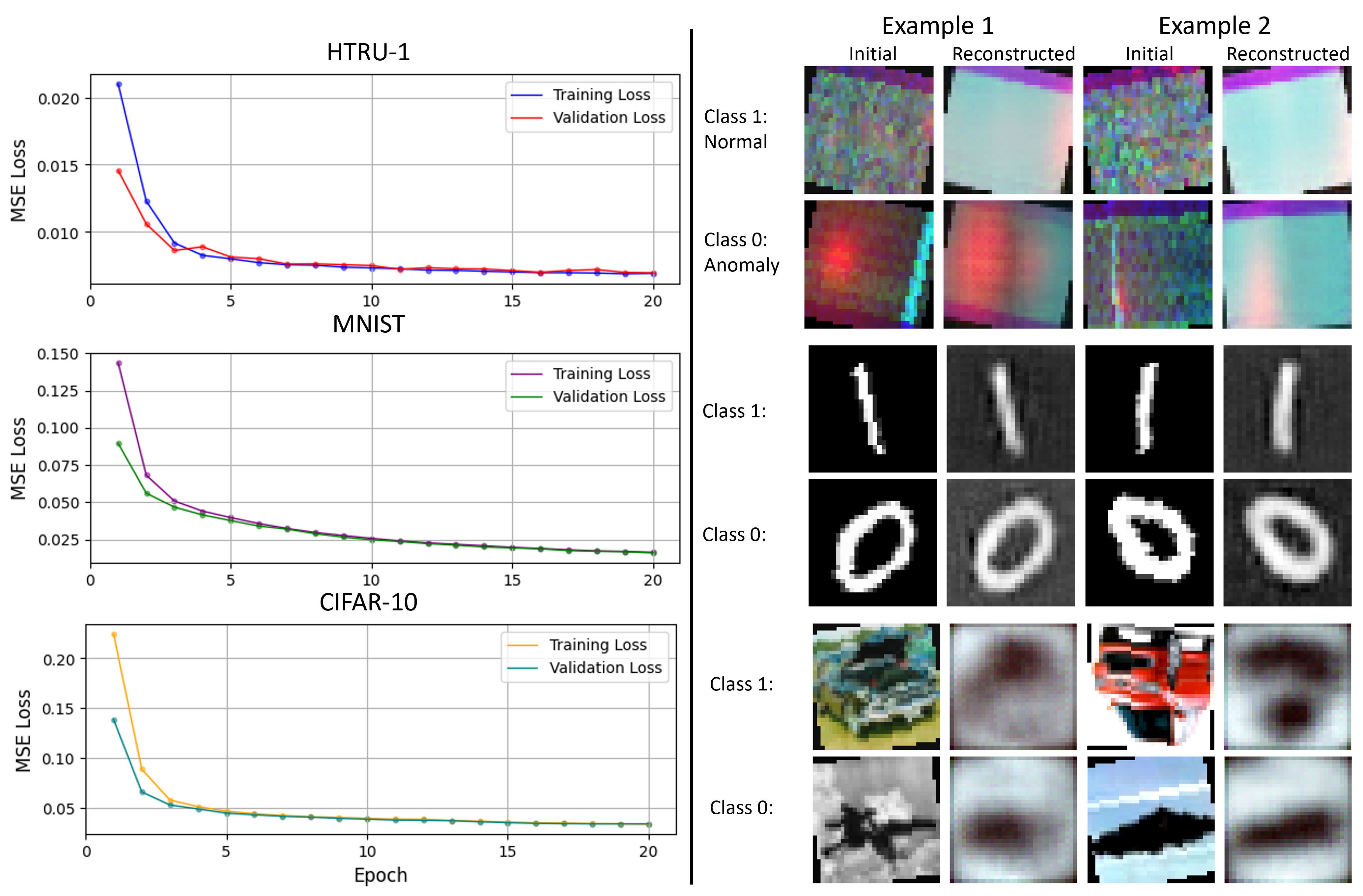}
  \caption{The loss curves for the HTRU-1, MNIST, and CIFAR-10 datasets. There is clear convergence observed in all three sets and the validation loss indicates limited overfitting. Two examples from each class of reconstructed images are included for reference.}
  \label{fig:figure3}
\end{figure*}

A natural question arises when comparing the performance of the same model architecture applied to different image datasets. MNIST consists of grayscale \(28 \times 28\) pixel images, whereas HTRU-1 and CIFAR-10 consist of three-channel \(32 \times 32\) pixel images. This leads to the concern that a similar model architecture, when applied to more complex images with additional channels, is expected to perform worse. A natural way to compensate for this would be to complicate the model architecture by changing the layers or increasing the size of the latent space. However, as previously mentioned, overly complex models, such as ResNet18, did not work well for HTRU-1. This raises the question: what about increasing the features in the latent space?

To investigate this, we utilized a simplified model architecture and applied it separately to the three grayscale channels corresponding to the three channels of the original HTRU-1 images. Thus, we have 64 features in the latent space per channel. If the results are similar, this indicates that, regardless of where the features are extracted from, the dataset does not lend itself to effective reconstruction and eventual separation. Conversely, if performance improves, it suggests that a larger latent space would be beneficial, however, it would also increase the qubit requirements in the quantum block.

In essence, we have three image reconstruction models, each yielding a separate set of 64 features for each channel. These features were used in their own respective sets of SVMs and OCSVMs (both classical and quantum). Overall, this approach involves three image reconstruction models and three pairs of SVMs for classification and anomaly detection. Additionally, we considered introducing weighted voting in an approach that resembles ensemble learning \cite{dietterich2000ensemble}. All hyperparameters were kept identical to those in the previous section.

\begin{figure*}[ht]
  \centering
  \includegraphics[width=\textwidth]{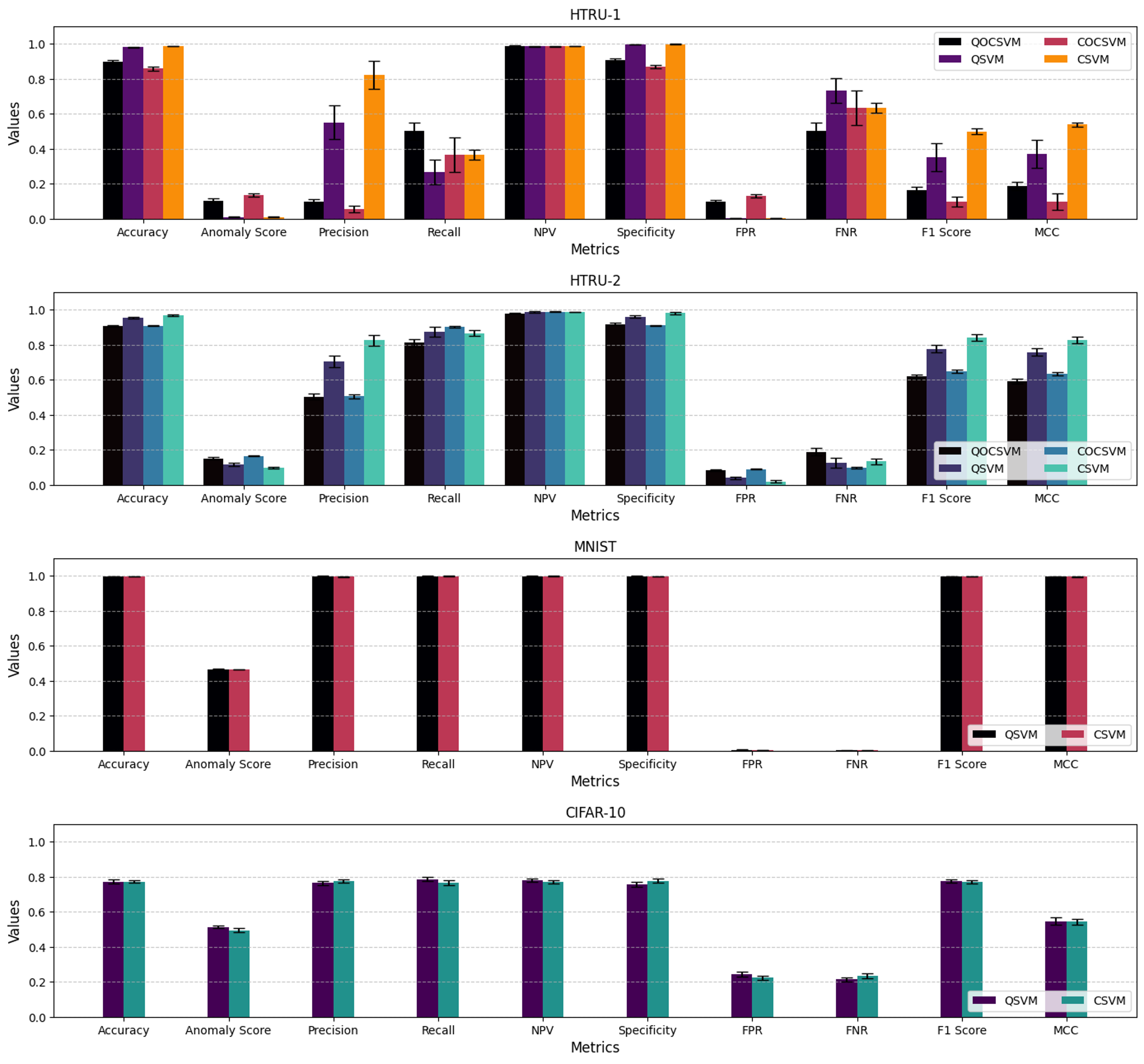}
  \caption{Performance metrics for the four chosen methods applied to the four datasets. Only standard two-class support vector machines are included for MNIST and CIFAR-10, since these sets are not unbalanced like the other two. Top to bottom: Performance metrics for HTRU-1, HTRU-2, MNIST, and CIFAR-10. Bars represent the average over three distinct runs and the error bars are the standard error.}
  \label{fig:figure5}
\end{figure*}

\section{\label{sec:Results} Results}

\subsection{Main Results}

\begin{figure}[b]
  \centering
  \includegraphics[width=\columnwidth]{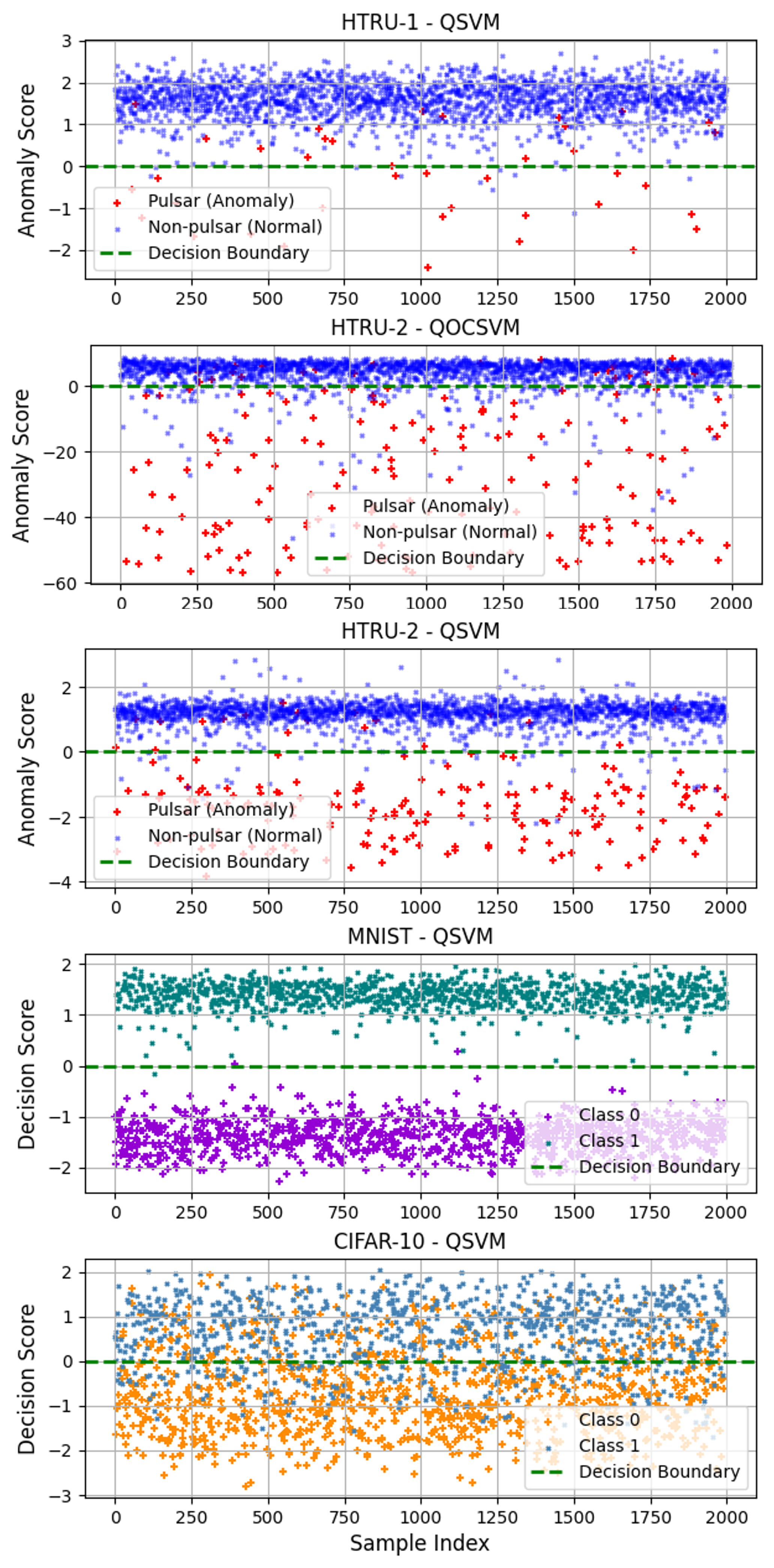}
  \caption{The decision functions from the SVMs and OCSVMs plotted as a horizontal line at zero represents the separating boundary in feature space. Above the line is the normal class and below the line the anomalies. Sample decision function values, called anomaly score (or decision score for MNIST and CIFAR-10) are plotted against sample index for visualization purposes. Top to bottom: QSVM for HTRU-1, QOCSVM and QSVM for HTRU-2, QSVM for MNIST, and finally QSVM for CIFAR-10. Only the one plot for HTRU-2 is an example of a QOCSVM boundary, since the QOCSVM boundary for HTRU-1 was random. All other three are QSVM boundaries and scores.}
  \label{fig:figure4}
\end{figure}

Using the autoencoder to encode the images into a reduced latent space of 64 features per sample, which is then amplitude encoded into 6 qubits for use in a QSVM and QOCSVM, along with classical results, leads to the results as shown in Figures \ref{fig:figure3}, \ref{fig:figure4} and \ref{fig:figure5}.

Before we get to the quantum machine learning results, we can look at the classical image reconstruction part. In Figure \ref{fig:figure3} we have plots that show the loss convergence of both the training and validation loss for the HTRU-1, MNIST, and CIFAR-10 datasets. In all cases the loss converges, with a convergent validation loss. This indicates limited overfitting in the image reconstruction process \cite{goodfellow2016deep}.

The images included in this figure are two examples of an initial image and its reconstructed image during testing, from each of the two classes in the datasets considered. Here it should visually be clear that the reconstruction for MNIST is nearly perfect, apart from slightly worse resolution. The HTRU-1 and CIFAR-10 images are reconstructed markedly worse. The final loss values, after convergence, are considerably low for both datasets, which indicates that either the model architecture is not complex enough for these datasets or that the datasets do not lend themselves well to being reproduced. From our testing with different architectures, it is clear that it is difficult to reproduce images as abstract as HTRU-1 images without a considerably deep autoencoder or a much larger dataset with more samples. Overall the model performs better for CIFAR-10 reconstruction and there is an indication of further improvement if we allowed for more epochs.

Figure \ref{fig:figure4} shows plots for the anomaly score (or decision score for the MNIST and CIFAR-10 cases) for some of the methods performed. The green horizontal line indicates the decision function at zero, which is a representation of the hyperplane in feature space. Any sample above the line, or inside the separating boundary, is considered normal and any sample below the line, or outisde the boundary, is considered anomalous. The $x$-axis is simply the sample index in the test set.

It is visually clear from the HTRU-2 plots that both the QSVM and QOCSVM are able to separately the data. There are some false predictions, but overall it seems to capture samples that are considered more anomalous than others. The HTRU-1 plot shown is an example of only the QSVM, which does not perform as well as for the HTRU-2 runs. Most anomalies are still flagged, but the proportion of mistakes is visually more pronounced. In the MNIST plot it is clear that the separation is really obvious, which connects well with the performance during reconstruction. CIFAR-10 performs worse, with a lot of sample overlap, but there is still a visual distinction. In this figure only the QSVM is given for MNIST and CIFAR-10, since it does not make sense to use a OCSVM for balanced datasets.

We can calculate the performance metrics of all the runs performed by taking the true positives, true negatives etc. from the confusion matrices or the anomaly score plots. These values are illustrated as a collection of histograms for comparison in Figure \ref{fig:figure5}.

The MNIST case clearly shows that the hybrid model of classically trained encoder and quantum block delivers essentially perfect results. This indicates that the features extracted by the encoder part, trained through image reconstruction, captures the necessary information for the machine learning task of classification through a SVM. CIFAR-10, just as the image reconstruction performed worse, the final classification performs worse as well. Here the expectation is precision and recall around $0.75$.

Applying the same for the HTRU-1 set is not as successful. Here the accuracy, NPV and specificity values are inflated, because of the imbalanced nature of the dataset. It is easier to predict negative samples correctly, even when randomly guessing. The two metrics to look at more carefully, as stated before, are the precision and recall metrics. The precision bars show that both the quantum and classical OCSVMs perform much worse as compared to the standard SVMs. Interestingly the quantum runs (QOCSVM) seems to perform slightly better than the classsical radial basis function SVMs. When we switch to a standard SVM, the classical SVM clearly performs better, with some variability in the metric indicated by error bars.

The recall bars indicate a more even performance between the four methods. The QOCSVM seems to perform the best out of the three methods, which is promising for its application as a quantum method for anomaly detection of imbalanced image datasets sent through encoders for feature extraction, but only if the image reconstruction is successful. Despite the improved performance over the other methods, the metric is still less than 0.6, which does not bode well for the pipeline as useful when using recall as a metric, specifically in this case for the HTRU-1 dataset.

The graph for HTRU-2, which serves as a benchmark of a simple dataset with only eight correlated features, shows what can typically be expected when applying only the quantum block. This is because the dataset in this case does not consist of images, implying that only the quantum block has to be applied here. The same pattern is observed in the precision, the main difference is that the metric values are simply higher. A promising observation is the recall performance here, which are all close to or above 0.8.

Another observation is that the false negative rate (FNR) for the HTRU-1 dataset is significantly higher compared to the other datasets. This aligns with the model's performance in predicting positive cases, particularly when evaluating precision and recall. Additionally, the false positive rate (FPR) is considerably lower, suggesting that the abundance of normal samples facilitates the model's ability to accurately identify these instances. This finding is further supported by the negative predictive value (NPV) and specificity metrics presented in all three graphs. The F1 score and Matthews correlation coefficient (MCC), which aim to balance these case-specific metrics, are also included in the analysis.

We include PPP only to guage the weighting applied during the SVMs and the 'nu' parameter during the OCSVMs. We told the models to only output anomalies at a rate corresponding to the original distribution so we would expect a value of around $0.5$ for MNIST and CIFAR-10, $0.09$ for HTRU-2 and $0.02$ for HTRU-1.

\subsection{Channel-wise approach}

\begin{figure}[ht]
  \centering
  \includegraphics[width=\columnwidth]{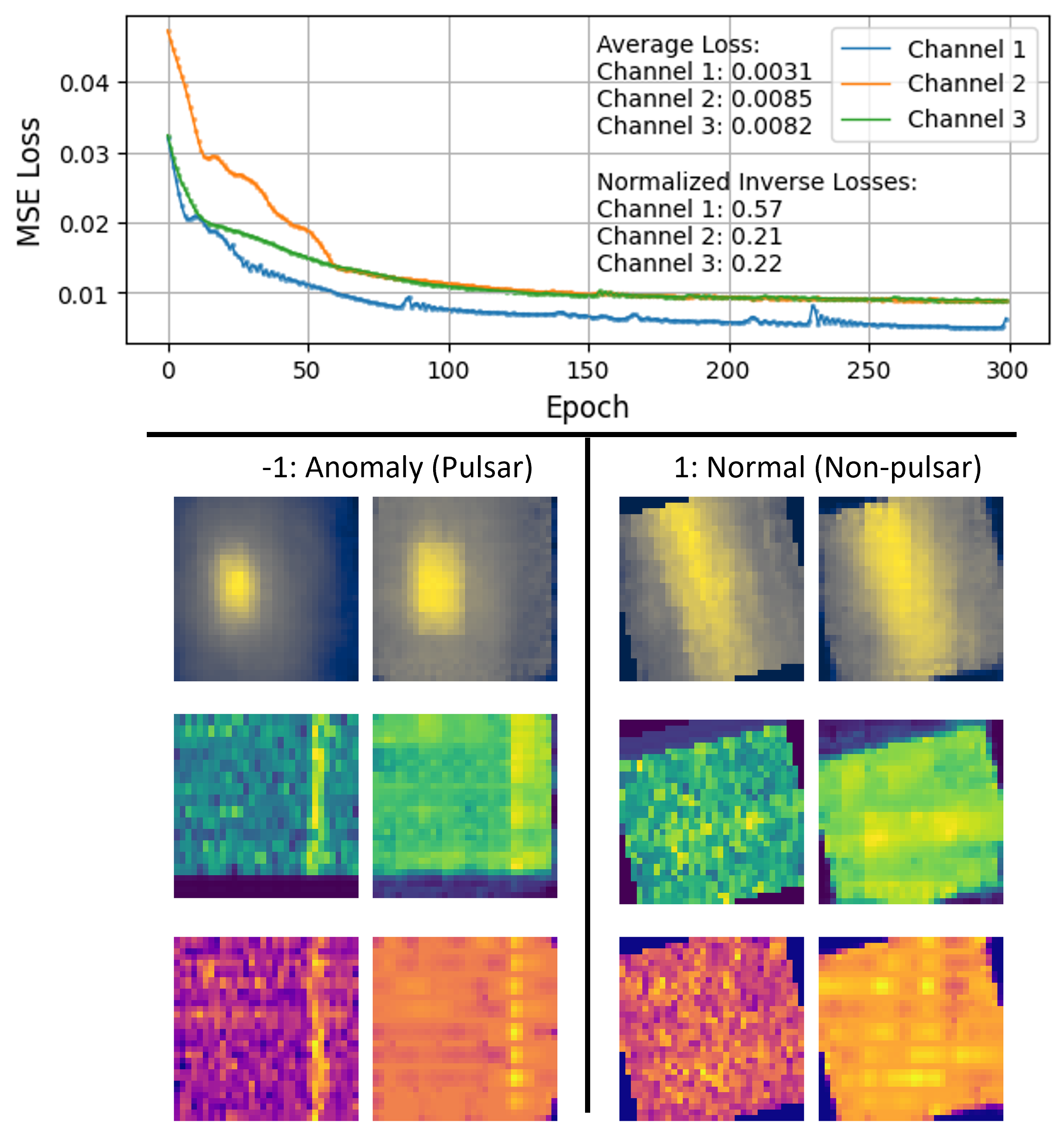}
  \caption{The loss curves for the channel-wise approach. Each channel of the HTRU-1 dataset is treated as gray scale images and features are extracted separately and fed into separate, but identical models. Some examples of reconstructed images are included. Top to bottom: Channel 1, 2, and then 3.}
  \label{fig:figure6}
\end{figure}

This section presents the results from treating each of the three channels in the HTRU-1 dataset as separate grayscale images, employing three identical models in a channel-wise approach. Based off the normalized inverse losses in Figure \ref{fig:figure6} the image reconstruction quality is best for channel 1. We also see that the simplified model architecture takes a lot more epochs to converge than our ResNet10 architecture. It also includes some examples of images that were reconstructed by the three separate models. These examples are visually consistent with the final normalized inverse losses.

Since channels 2 and 3 performed worse, the expected consequence would be an improved classification and anomaly detection performance for the best performing image reconstruction model, but instead the better performance was observed for channels 2 and 3. An explanation for this could be the fact that channel 1 images typically contain a concentration of intensity that is easier to reconstruct than the occasional intense vertical lines in channels 2 and 3 for the anomalous samples. It might be easier for the SVMs to distinguish between samples that have no vertical lines and ones that have traces of vertical lines in sub-optimally reconstructed images, than in images that are reconstructed at a better quality, but with no obvious distinguishing features. The suggestion would be to implement a weighted voting scheme in an ensemble learning machine learning approach if the decision is to treat images as three separate gray scale images. How the weighting of this ensemble approach would be chosen is unclear. Here we considered using the inverse losses, however, as stated earlier, this would not make much sense, since the best performing reconstruction model did not yield the best performing anomaly detection model.

\begin{figure*}[ht]
  \centering
  \includegraphics[width=\textwidth]{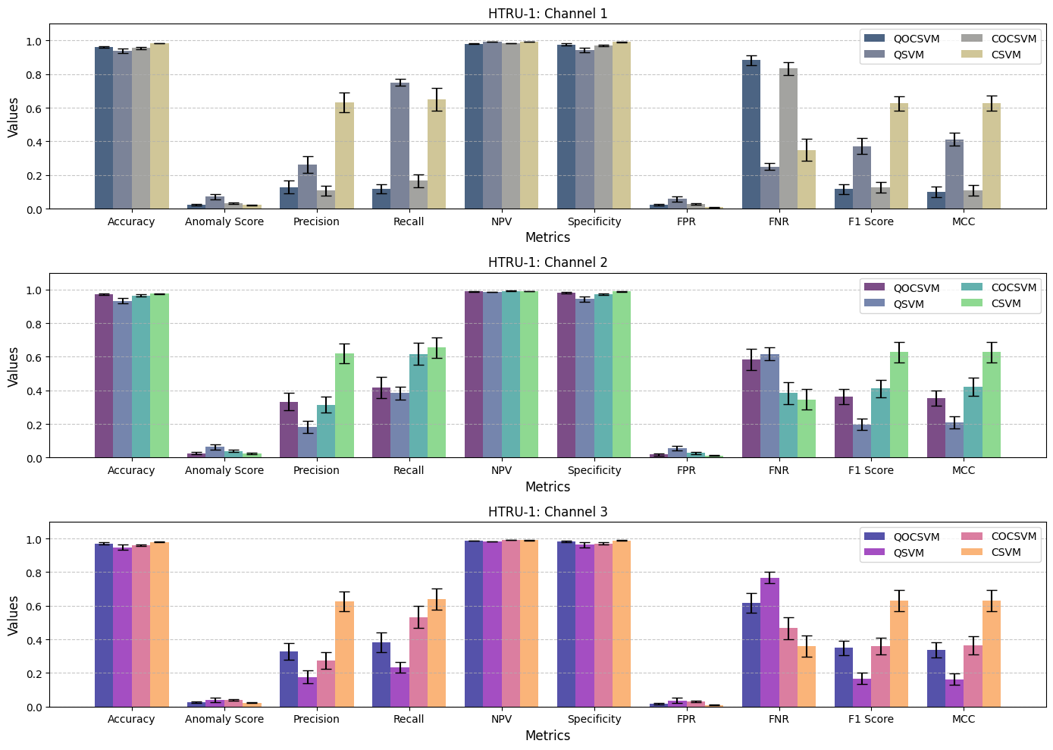}
  \caption{Performance metrics for the channel-wise approach. Each set of bars are the performance metrics for only one of the channels in the HTRU-1 dataset images. Bars represent the average of six distinct runs and the error bars are the standard error.}
  \label{fig:figure7}
\end{figure*}

Comparing the channel-wise approach histograms  in Figure \ref{fig:figure7} to the main results histograms from before, the observations are more complicated. Here the recall and precision values will obviously depend on which channel is being used in the pipeline. Some channels will contain more information that is useful for the SVMs to leverage for classification. Here the indication is that precision in the QOCSVM is generally lower for channel 1 as compared to channel 2 and 3, whereas the QSVM performs better for channel 1. Channels 2 and 3 are seemingly better for anomaly detection and channel 1 is better for classification. In terms of recall, a different observation is made. The results are still channel dependent, but now the QSVM outperforms all other methods, including the CSVM for channel 1. The classical methods outperform the quantum methods for channel 2 and 3, however the QOCSVM wins against the COCSVM, while the CSVM wins out against the QSVM. In this case you would prefer to do anomaly detection with a quantum model and classification with a classical model. The other metrics tell a similar story as was observed in the main results section.

\section{\label{sec:Conclusion} Conclusion}

We used a classical ResNet10-inspired convolutional autoencoder (CAE) trained for image reconstruction to extract 64 abstract features in the latent space of the encoder for input into simulations of a quantum machine learning block made up of either a QSVM or QOCSVM. We applied this to MNIST, CIFAR-10 and the HTRU-1 image datasets. We also applied the quantum blocks to the HTRU-2 dataset, which contains correlated features to serve as benchmark. This was all compared against classical counterparts. Also included is a channel-wise approach, while using a simpler architecture that outputs 64 features in the latent space per channel.

The feature extractor trained for image reconstruction is able to extract meaningful features for classification through our classical and quantum SVMs for MNIST. The performance of the hybrid pipeline does not perform optimally for the HTRU-1 image set, which can be attributed to a few reasons, namely the abstractness of the images, the information contained within the image itself, the lack sufficient model complexity for the HTRU-1 images etc. This was addressed in part by introducing a channel-wise approach, which essentially tripled the amount of features that were extracted from the input images. Despite the increase in the number of features extracted, which essentially meant an increase in qubit number, it did not dramatically improve the model. We observed channel-dependent results, where channels 2 and 3 tend to be more important for anomaly detection. The explanation for this was that channels 2 and 3 contained vertical lines that are easier to distinguish for a OCSVM, despite channel 1's better image reconstruction quality. Overall the HTRU-1 set proved to be too abstract for an image reconstruction task and ultimately leads to less than optimal results during separation. The CIFAR-10 results fall directly in the middle between MNIST and HTRU-1. It peforms at an expected recall and precision of around 0.75, which indicates separation, despite visually poor image reconstruction. The image reconstruction could, however, be improved if more epochs were allowed. Finally, the HTRU-2 benchmark shows that anomaly detection is possible for a dataset consisting of well-correlated features and is viewed as another middle point between the perfect MNIST results and what the HTRU-1 images.

We believe that the classical image reconstruction pipeline, coupled with a quantum block can be implemented for image classification and anomaly detection, dependent on the image set in question. This is obvious from the holistically different results for MNIST, CIFAR-10, and HTRU-1.

This work can be extended by performing a broad architecture search to find the optimal autoencoder for image reconstruction and to try out other quantum blocks, such as variational circuits \cite{wang2023quantum, senokosov2024quantum}. Other feature extraction methods can also be looked at, where examples of some ideas we considered included FFT, edge detection, and noise removal. Running the pipeline through real hardware would be the main missing link, however, the expected results from this study would be far worse if they were to be run on real quantum devices. Quantum machine learning in general is severely limited by the current NISQ-era hardware, not only because of noise, but also because of runtime requirements.

\begin{acknowledgments}
We wish to thank Amy Rouillard for discussions related to autoencoders and image reconstruction tasks.
\end{acknowledgments}

\section*{Funding}

This work was funded by the South African Quantum Technology Initiative (SA QuTI) through the Department of Science and Innovation of South Africa.

\section*{Data Availability Statement}
\label{sec:data_availability}

The \href{https://as595.github.io/HTRU1/}{HTRU-1}, \href{https://archive.ics.uci.edu/dataset/372/htru2}{HTRU-2}, \href{http://yann.lecun.com/exdb/mnist/}{MNIST}, and \href{https://www.cs.toronto.edu/~kriz/cifar.html}{CIFAR-10} are available for download. MNIST and CIFAR-10 can also be imported from machine learning libraries such as TensorFlow and PyTorch.

\begin{table*}[t]
\centering
\begin{tabular}{lcccc}
\hline
\textbf{Section} & \textbf{Layer} & \textbf{Details} & \textbf{Input size} & \textbf{Output size} \\ \hline
\multirow{6}{*}{\textbf{Encoder}} & \texttt{Layer1} & Conv2d + BatchNorm2d + MaxPool2d & (Channels, H, W) & (64, H/4, W/4) \\ 
& \texttt{Layer2} & Conv2d + BatchNorm2d & (64, H/4, W/4) & (128, H/8, W/8) \\ 
& \texttt{Layer3} & Conv2d + BatchNorm2d & (128, H/8, W/8) & (256, H/16, W/16) \\ 
& \texttt{Layer4} & Conv2d + BatchNorm2d & (256, H/16, W/16) & (64, H/32, W/32) \\ 
& \texttt{Avgpool} & AdaptiveAvgPool2d & (64, H/32, W/32) & (64, 1, 1) \\ 
& \texttt{Flatten} & - & (64, 1, 1) & (64) \\ \hline
\multirow{5}{*}{\textbf{Decoder}} & \texttt{Layer1} & ConvTranspose2d & (64, 1, 1) & (32, 4, 4) \\ 
& \texttt{Layer2} & ConvTranspose2d & (32, 4, 4) & (16, 8, 8) \\ 
& \texttt{Layer3} & ConvTranspose2d & (16, 8, 8) & (8, 16, 16) \\ 
& \texttt{Layer4} & ConvTranspose2d & (8, 16, 16) & (4, 32, 32) \\ 
& \texttt{Layer5} & ConvTranspose2d & (4, 32, 32) & (Channels, H, W) \\ \hline
\end{tabular}
\caption{Summary of the overall CAE architecture. A stride of 2, padding of 1 and a kernel size of 3 was used throughout the encoder. In the first layer of the encoder the kernel size was changed to 7 and the padding to 3. The decoder had a consistent kernel size of 4, padding of 4 and a stride of 2 throughout its architecture. The activation functions after every layer were ReLU functions apart from right after the final decoder layer where Tanh was used instead. We crop the final output of the decoder to the size of the reconstructed image for matching in the loss calculations, which implies that some information is lost when doing this for MNIST. The most important part to note is that after the avgpool layer and flattening it, the total amount of abstract features were 64. We specifically changed the parameters for this to be the case.}
\label{table:encoder_decoder}
\end{table*}

\appendix

\section{Performance Metrics}
\label{sec:performance_metrics}

Accuracy measures the proportion of correctly classified samples (both normal and anomalous) out of the total number of samples, giving an overall indication of the model's performance. Negative Prediction Value (NPV) assesses the proportion of correctly predicted normal cases from all samples predicted as normal, reflecting how well the model identifies normal cases. Specificity quantifies the proportion of correctly predicted normal cases out of all normal cases, indicating the model's ability to correctly identify non-anomalies. The False Positive Rate (FPR) represents the proportion of normal cases incorrectly predicted as anomalies, highlighting the rate of false alarms. The False Negative Rate (FNR) shows the proportion of anomalies incorrectly predicted as normal cases, pointing out how often anomalies are missed. The F1 Score combines precision and recall into a single metric, providing a balanced measure of the model's performance on positive cases. Matthew's Correlation Coefficient (MCC) offers a comprehensive measure of binary classification quality by considering all four confusion matrix categories, particularly useful for imbalanced datasets.

In imbalanced datasets, accuracy, Negative Prediction Value (NPV), and specificity can be misleading. Accuracy may be high if the model predominantly predicts the majority class, failing to capture the minority class accurately. NPV can be inflated if the dataset has many more normal cases, making it seem like the model is performing well in predicting normal cases, even if it misses anomalies. Specificity might also appear high if the majority class is dominant, giving a false impression of model performance on the minority class. Metrics focusing on the minority class, like precision and recall, often provide a clearer picture of model effectiveness in such cases, which is why particular emphasis is placed on these metrics.

\section{Autoencoder Architecture}
\label{sec:Autoencoder Architecture}

There are three convolutional autoencoder architectures to take note of, all of which result in 64 features in the latent space. The main architecture for MNIST, which takes as input gray scale 28x28 pixel images, the main architecture for HTRU-1 and CIFAR-10, which take as input three-channel 32x32 pixel images, and the three separate channel-wise models which take as input single-channel 32x32 pixel images as input. The two main architectures are given in Table \ref{table:encoder_decoder}, where it is important to note that the final output layer is cropped to either 28x28 or 32x32 depending on the dataset before matching in the mean squared error loss calculation. 

The channel-wise approach autoencoder architecture is designed to compress and reconstruct grayscale images using a simplified architecture. The encoder consists of three convolutional layers, each with a kernel size of 3x3, strides of 2, and padding of 1, reducing the spatial dimensions and channel size from 16 to 8 to 4. ReLU activations were utilized. The decoder employs three transposed convolutional layers with the same kernel size, strides, and padding. An output padding of 1 to reverse the spatial reduction was used. The decoder works by progressively increasing the channel size from 4 to 8 to 16. The final layer restores the original input channel count, aiming to accurately reconstruct the input images. 

\bibliography{main}

\begin{thebibliography}{81}%
\makeatletter
\providecommand \@ifxundefined [1]{%
 \@ifx{#1\undefined}
}%
\providecommand \@ifnum [1]{%
 \ifnum #1\expandafter \@firstoftwo
 \else \expandafter \@secondoftwo
 \fi
}%
\providecommand \@ifx [1]{%
 \ifx #1\expandafter \@firstoftwo
 \else \expandafter \@secondoftwo
 \fi
}%
\providecommand \natexlab [1]{#1}%
\providecommand \enquote  [1]{``#1''}%
\providecommand \bibnamefont  [1]{#1}%
\providecommand \bibfnamefont [1]{#1}%
\providecommand \citenamefont [1]{#1}%
\providecommand \href@noop [0]{\@secondoftwo}%
\providecommand \href [0]{\begingroup \@sanitize@url \@href}%
\providecommand \@href[1]{\@@startlink{#1}\@@href}%
\providecommand \@@href[1]{\endgroup#1\@@endlink}%
\providecommand \@sanitize@url [0]{\catcode `\\12\catcode `\$12\catcode `\&12\catcode `\#12\catcode `\^12\catcode `\_12\catcode `\%12\relax}%
\providecommand \@@startlink[1]{}%
\providecommand \@@endlink[0]{}%
\providecommand \url  [0]{\begingroup\@sanitize@url \@url }%
\providecommand \@url [1]{\endgroup\@href {#1}{\urlprefix }}%
\providecommand \urlprefix  [0]{URL }%
\providecommand \Eprint [0]{\href }%
\providecommand \doibase [0]{https://doi.org/}%
\providecommand \selectlanguage [0]{\@gobble}%
\providecommand \bibinfo  [0]{\@secondoftwo}%
\providecommand \bibfield  [0]{\@secondoftwo}%
\providecommand \translation [1]{[#1]}%
\providecommand \BibitemOpen [0]{}%
\providecommand \bibitemStop [0]{}%
\providecommand \bibitemNoStop [0]{.\EOS\space}%
\providecommand \EOS [0]{\spacefactor3000\relax}%
\providecommand \BibitemShut  [1]{\csname bibitem#1\endcsname}%
\let\auto@bib@innerbib\@empty
\bibitem [{\citenamefont {Biamonte}\ \emph {et~al.}(2017)\citenamefont {Biamonte}, \citenamefont {Wittek}, \citenamefont {Pancotti}, \citenamefont {Rebentrost}, \citenamefont {Wiebe},\ and\ \citenamefont {Lloyd}}]{Biamonte2017}%
  \BibitemOpen
  \bibfield  {author} {\bibinfo {author} {\bibfnamefont {J.}~\bibnamefont {Biamonte}}, \bibinfo {author} {\bibfnamefont {P.}~\bibnamefont {Wittek}}, \bibinfo {author} {\bibfnamefont {N.}~\bibnamefont {Pancotti}}, \bibinfo {author} {\bibfnamefont {P.}~\bibnamefont {Rebentrost}}, \bibinfo {author} {\bibfnamefont {N.}~\bibnamefont {Wiebe}},\ and\ \bibinfo {author} {\bibfnamefont {S.}~\bibnamefont {Lloyd}},\ }\bibfield  {title} {\bibinfo {title} {Quantum machine learning},\ }\href {https://doi.org/10.1038/nature23474} {\bibfield  {journal} {\bibinfo  {journal} {Nature}\ }\textbf {\bibinfo {volume} {549}},\ \bibinfo {pages} {195} (\bibinfo {year} {2017})}\BibitemShut {NoStop}%
\bibitem [{\citenamefont {Schuld}\ \emph {et~al.}(2015)\citenamefont {Schuld}, \citenamefont {Sinayskiy},\ and\ \citenamefont {Petruccione}}]{schuld2015introduction}%
  \BibitemOpen
  \bibfield  {author} {\bibinfo {author} {\bibfnamefont {M.}~\bibnamefont {Schuld}}, \bibinfo {author} {\bibfnamefont {I.}~\bibnamefont {Sinayskiy}},\ and\ \bibinfo {author} {\bibfnamefont {F.}~\bibnamefont {Petruccione}},\ }\bibfield  {title} {\bibinfo {title} {An introduction to quantum machine learning},\ }\href@noop {} {\bibfield  {journal} {\bibinfo  {journal} {Contemporary Physics}\ }\textbf {\bibinfo {volume} {56}},\ \bibinfo {pages} {172} (\bibinfo {year} {2015})}\BibitemShut {NoStop}%
\bibitem [{\citenamefont {Schuld}\ and\ \citenamefont {Petruccione}(2021)}]{schuld2021machine}%
  \BibitemOpen
  \bibfield  {author} {\bibinfo {author} {\bibfnamefont {M.}~\bibnamefont {Schuld}}\ and\ \bibinfo {author} {\bibfnamefont {F.}~\bibnamefont {Petruccione}},\ }\href {https://doi.org/10.1007/978-3-030-83098-4} {\emph {\bibinfo {title} {Machine Learning with Quantum Computers}}},\ \bibinfo {edition} {2nd}\ ed.,\ Quantum Science and Technology\ (\bibinfo  {publisher} {Springer Cham},\ \bibinfo {year} {2021})\BibitemShut {NoStop}%
\bibitem [{\citenamefont {Nielsen}\ and\ \citenamefont {Chuang}(2010)}]{nielsen2010quantum}%
  \BibitemOpen
  \bibfield  {author} {\bibinfo {author} {\bibfnamefont {M.~A.}\ \bibnamefont {Nielsen}}\ and\ \bibinfo {author} {\bibfnamefont {I.~L.}\ \bibnamefont {Chuang}},\ }\href@noop {} {\emph {\bibinfo {title} {Quantum computation and quantum information}}}\ (\bibinfo  {publisher} {Cambridge university press},\ \bibinfo {year} {2010})\BibitemShut {NoStop}%
\bibitem [{\citenamefont {Bharti}\ \emph {et~al.}(2022)\citenamefont {Bharti}, \citenamefont {Cervera-Lierta}, \citenamefont {Kyaw}, \citenamefont {Haug}, \citenamefont {Alperin-Lea}, \citenamefont {Anand}, \citenamefont {Degroote}, \citenamefont {Heimonen}, \citenamefont {Kottmann}, \citenamefont {Menke} \emph {et~al.}}]{bharti2022noisy}%
  \BibitemOpen
  \bibfield  {author} {\bibinfo {author} {\bibfnamefont {K.}~\bibnamefont {Bharti}}, \bibinfo {author} {\bibfnamefont {A.}~\bibnamefont {Cervera-Lierta}}, \bibinfo {author} {\bibfnamefont {T.~H.}\ \bibnamefont {Kyaw}}, \bibinfo {author} {\bibfnamefont {T.}~\bibnamefont {Haug}}, \bibinfo {author} {\bibfnamefont {S.}~\bibnamefont {Alperin-Lea}}, \bibinfo {author} {\bibfnamefont {A.}~\bibnamefont {Anand}}, \bibinfo {author} {\bibfnamefont {M.}~\bibnamefont {Degroote}}, \bibinfo {author} {\bibfnamefont {H.}~\bibnamefont {Heimonen}}, \bibinfo {author} {\bibfnamefont {J.~S.}\ \bibnamefont {Kottmann}}, \bibinfo {author} {\bibfnamefont {T.}~\bibnamefont {Menke}}, \emph {et~al.},\ }\bibfield  {title} {\bibinfo {title} {Noisy intermediate-scale quantum algorithms},\ }\href@noop {} {\bibfield  {journal} {\bibinfo  {journal} {Reviews of Modern Physics}\ }\textbf {\bibinfo {volume} {94}},\ \bibinfo {pages} {015004} (\bibinfo {year} {2022})}\BibitemShut {NoStop}%
\bibitem [{\citenamefont {Cheng}\ \emph {et~al.}(2023)\citenamefont {Cheng}, \citenamefont {Deng}, \citenamefont {Gu}, \citenamefont {He}, \citenamefont {Hu}, \citenamefont {Huang}, \citenamefont {Li}, \citenamefont {Lin}, \citenamefont {Lu}, \citenamefont {Lu} \emph {et~al.}}]{cheng2023noisy}%
  \BibitemOpen
  \bibfield  {author} {\bibinfo {author} {\bibfnamefont {B.}~\bibnamefont {Cheng}}, \bibinfo {author} {\bibfnamefont {X.-H.}\ \bibnamefont {Deng}}, \bibinfo {author} {\bibfnamefont {X.}~\bibnamefont {Gu}}, \bibinfo {author} {\bibfnamefont {Y.}~\bibnamefont {He}}, \bibinfo {author} {\bibfnamefont {G.}~\bibnamefont {Hu}}, \bibinfo {author} {\bibfnamefont {P.}~\bibnamefont {Huang}}, \bibinfo {author} {\bibfnamefont {J.}~\bibnamefont {Li}}, \bibinfo {author} {\bibfnamefont {B.-C.}\ \bibnamefont {Lin}}, \bibinfo {author} {\bibfnamefont {D.}~\bibnamefont {Lu}}, \bibinfo {author} {\bibfnamefont {Y.}~\bibnamefont {Lu}}, \emph {et~al.},\ }\bibfield  {title} {\bibinfo {title} {Noisy intermediate-scale quantum computers},\ }\href@noop {} {\bibfield  {journal} {\bibinfo  {journal} {Frontiers of Physics}\ }\textbf {\bibinfo {volume} {18}},\ \bibinfo {pages} {21308} (\bibinfo {year} {2023})}\BibitemShut {NoStop}%
\bibitem [{\citenamefont {Preskill}(2018)}]{preskill2018quantum}%
  \BibitemOpen
  \bibfield  {author} {\bibinfo {author} {\bibfnamefont {J.}~\bibnamefont {Preskill}},\ }\bibfield  {title} {\bibinfo {title} {Quantum computing in the nisq era and beyond},\ }\href@noop {} {\bibfield  {journal} {\bibinfo  {journal} {Quantum}\ }\textbf {\bibinfo {volume} {2}},\ \bibinfo {pages} {79} (\bibinfo {year} {2018})}\BibitemShut {NoStop}%
\bibitem [{\citenamefont {Chen}\ \emph {et~al.}(2023)\citenamefont {Chen}, \citenamefont {Cotler}, \citenamefont {Huang},\ and\ \citenamefont {Li}}]{chen2023complexity}%
  \BibitemOpen
  \bibfield  {author} {\bibinfo {author} {\bibfnamefont {S.}~\bibnamefont {Chen}}, \bibinfo {author} {\bibfnamefont {J.}~\bibnamefont {Cotler}}, \bibinfo {author} {\bibfnamefont {H.-Y.}\ \bibnamefont {Huang}},\ and\ \bibinfo {author} {\bibfnamefont {J.}~\bibnamefont {Li}},\ }\bibfield  {title} {\bibinfo {title} {The complexity of nisq},\ }\href@noop {} {\bibfield  {journal} {\bibinfo  {journal} {Nature Communications}\ }\textbf {\bibinfo {volume} {14}},\ \bibinfo {pages} {6001} (\bibinfo {year} {2023})}\BibitemShut {NoStop}%
\bibitem [{\citenamefont {Shor}(1996)}]{shor1996fault}%
  \BibitemOpen
  \bibfield  {author} {\bibinfo {author} {\bibfnamefont {P.~W.}\ \bibnamefont {Shor}},\ }\bibfield  {title} {\bibinfo {title} {Fault-tolerant quantum computation},\ }in\ \href@noop {} {\emph {\bibinfo {booktitle} {Proceedings of 37th conference on foundations of computer science}}}\ (\bibinfo {organization} {IEEE},\ \bibinfo {year} {1996})\ pp.\ \bibinfo {pages} {56--65}\BibitemShut {NoStop}%
\bibitem [{\citenamefont {Preskill}(1998)}]{preskill1998fault}%
  \BibitemOpen
  \bibfield  {author} {\bibinfo {author} {\bibfnamefont {J.}~\bibnamefont {Preskill}},\ }\bibfield  {title} {\bibinfo {title} {Fault-tolerant quantum computation},\ }in\ \href@noop {} {\emph {\bibinfo {booktitle} {Introduction to quantum computation and information}}}\ (\bibinfo  {publisher} {World Scientific},\ \bibinfo {year} {1998})\ pp.\ \bibinfo {pages} {213--269}\BibitemShut {NoStop}%
\bibitem [{\citenamefont {DiVincenzo}(2000)}]{divincenzo2000physical}%
  \BibitemOpen
  \bibfield  {author} {\bibinfo {author} {\bibfnamefont {D.~P.}\ \bibnamefont {DiVincenzo}},\ }\bibfield  {title} {\bibinfo {title} {The physical implementation of quantum computation},\ }\href@noop {} {\bibfield  {journal} {\bibinfo  {journal} {Fortschritte der Physik: Progress of Physics}\ }\textbf {\bibinfo {volume} {48}},\ \bibinfo {pages} {771} (\bibinfo {year} {2000})}\BibitemShut {NoStop}%
\bibitem [{\citenamefont {Shor}(1995)}]{shor1995scheme}%
  \BibitemOpen
  \bibfield  {author} {\bibinfo {author} {\bibfnamefont {P.~W.}\ \bibnamefont {Shor}},\ }\bibfield  {title} {\bibinfo {title} {Scheme for reducing decoherence in quantum computer memory},\ }\href@noop {} {\bibfield  {journal} {\bibinfo  {journal} {Physical review A}\ }\textbf {\bibinfo {volume} {52}},\ \bibinfo {pages} {R2493} (\bibinfo {year} {1995})}\BibitemShut {NoStop}%
\bibitem [{\citenamefont {Steane}(1996)}]{steane1996error}%
  \BibitemOpen
  \bibfield  {author} {\bibinfo {author} {\bibfnamefont {A.~M.}\ \bibnamefont {Steane}},\ }\bibfield  {title} {\bibinfo {title} {Error correcting codes in quantum theory},\ }\href@noop {} {\bibfield  {journal} {\bibinfo  {journal} {Physical Review Letters}\ }\textbf {\bibinfo {volume} {77}},\ \bibinfo {pages} {793} (\bibinfo {year} {1996})}\BibitemShut {NoStop}%
\bibitem [{\citenamefont {Lidar}\ and\ \citenamefont {Brun}(2013)}]{lidar2013quantum}%
  \BibitemOpen
  \bibfield  {author} {\bibinfo {author} {\bibfnamefont {D.~A.}\ \bibnamefont {Lidar}}\ and\ \bibinfo {author} {\bibfnamefont {T.~A.}\ \bibnamefont {Brun}},\ }\href@noop {} {\emph {\bibinfo {title} {Quantum error correction}}}\ (\bibinfo  {publisher} {Cambridge university press},\ \bibinfo {year} {2013})\BibitemShut {NoStop}%
\bibitem [{\citenamefont {Knill}\ and\ \citenamefont {Laflamme}(1997)}]{knill1997theory}%
  \BibitemOpen
  \bibfield  {author} {\bibinfo {author} {\bibfnamefont {E.}~\bibnamefont {Knill}}\ and\ \bibinfo {author} {\bibfnamefont {R.}~\bibnamefont {Laflamme}},\ }\bibfield  {title} {\bibinfo {title} {Theory of quantum error-correcting codes},\ }\href@noop {} {\bibfield  {journal} {\bibinfo  {journal} {Physical Review A}\ }\textbf {\bibinfo {volume} {55}},\ \bibinfo {pages} {900} (\bibinfo {year} {1997})}\BibitemShut {NoStop}%
\bibitem [{\citenamefont {Reddy}\ \emph {et~al.}(2020)\citenamefont {Reddy}, \citenamefont {Reddy}, \citenamefont {Lakshmanna}, \citenamefont {Kaluri}, \citenamefont {Rajput}, \citenamefont {Srivastava},\ and\ \citenamefont {Baker}}]{reddy2020analysis}%
  \BibitemOpen
  \bibfield  {author} {\bibinfo {author} {\bibfnamefont {G.~T.}\ \bibnamefont {Reddy}}, \bibinfo {author} {\bibfnamefont {M.~P.~K.}\ \bibnamefont {Reddy}}, \bibinfo {author} {\bibfnamefont {K.}~\bibnamefont {Lakshmanna}}, \bibinfo {author} {\bibfnamefont {R.}~\bibnamefont {Kaluri}}, \bibinfo {author} {\bibfnamefont {D.~S.}\ \bibnamefont {Rajput}}, \bibinfo {author} {\bibfnamefont {G.}~\bibnamefont {Srivastava}},\ and\ \bibinfo {author} {\bibfnamefont {T.}~\bibnamefont {Baker}},\ }\bibfield  {title} {\bibinfo {title} {Analysis of dimensionality reduction techniques on big data},\ }\href@noop {} {\bibfield  {journal} {\bibinfo  {journal} {Ieee Access}\ }\textbf {\bibinfo {volume} {8}},\ \bibinfo {pages} {54776} (\bibinfo {year} {2020})}\BibitemShut {NoStop}%
\bibitem [{\citenamefont {Ghosh}(2021)}]{ghosh2021encoding}%
  \BibitemOpen
  \bibfield  {author} {\bibinfo {author} {\bibfnamefont {K.}~\bibnamefont {Ghosh}},\ }\bibfield  {title} {\bibinfo {title} {Encoding classical data into a quantum computer},\ }\href@noop {} {\bibfield  {journal} {\bibinfo  {journal} {arXiv preprint arXiv:2107.09155}\ } (\bibinfo {year} {2021})}\BibitemShut {NoStop}%
\bibitem [{\citenamefont {Wright}\ \emph {et~al.}(2010)\citenamefont {Wright}, \citenamefont {Eisenhardt}, \citenamefont {Mainzer}, \citenamefont {Ressler}, \citenamefont {Cutri}, \citenamefont {Jarrett}, \citenamefont {Kirkpatrick}, \citenamefont {Padgett}, \citenamefont {McMillan}, \citenamefont {Skrutskie} \emph {et~al.}}]{wright2010wide}%
  \BibitemOpen
  \bibfield  {author} {\bibinfo {author} {\bibfnamefont {E.~L.}\ \bibnamefont {Wright}}, \bibinfo {author} {\bibfnamefont {P.~R.}\ \bibnamefont {Eisenhardt}}, \bibinfo {author} {\bibfnamefont {A.~K.}\ \bibnamefont {Mainzer}}, \bibinfo {author} {\bibfnamefont {M.~E.}\ \bibnamefont {Ressler}}, \bibinfo {author} {\bibfnamefont {R.~M.}\ \bibnamefont {Cutri}}, \bibinfo {author} {\bibfnamefont {T.}~\bibnamefont {Jarrett}}, \bibinfo {author} {\bibfnamefont {J.~D.}\ \bibnamefont {Kirkpatrick}}, \bibinfo {author} {\bibfnamefont {D.}~\bibnamefont {Padgett}}, \bibinfo {author} {\bibfnamefont {R.~S.}\ \bibnamefont {McMillan}}, \bibinfo {author} {\bibfnamefont {M.}~\bibnamefont {Skrutskie}}, \emph {et~al.},\ }\bibfield  {title} {\bibinfo {title} {The wide-field infrared survey explorer (wise): mission description and initial on-orbit performance},\ }\href@noop {} {\bibfield  {journal} {\bibinfo  {journal} {The Astronomical Journal}\ }\textbf {\bibinfo {volume} {140}},\ \bibinfo {pages} {1868} (\bibinfo {year}
  {2010})}\BibitemShut {NoStop}%
\bibitem [{\citenamefont {York}\ \emph {et~al.}(2000)\citenamefont {York}, \citenamefont {Adelman}, \citenamefont {Anderson~Jr}, \citenamefont {Anderson}, \citenamefont {Annis}, \citenamefont {Bahcall}, \citenamefont {Bakken}, \citenamefont {Barkhouser}, \citenamefont {Bastian}, \citenamefont {Berman} \emph {et~al.}}]{york2000sloan}%
  \BibitemOpen
  \bibfield  {author} {\bibinfo {author} {\bibfnamefont {D.~G.}\ \bibnamefont {York}}, \bibinfo {author} {\bibfnamefont {J.}~\bibnamefont {Adelman}}, \bibinfo {author} {\bibfnamefont {J.~E.}\ \bibnamefont {Anderson~Jr}}, \bibinfo {author} {\bibfnamefont {S.~F.}\ \bibnamefont {Anderson}}, \bibinfo {author} {\bibfnamefont {J.}~\bibnamefont {Annis}}, \bibinfo {author} {\bibfnamefont {N.~A.}\ \bibnamefont {Bahcall}}, \bibinfo {author} {\bibfnamefont {J.}~\bibnamefont {Bakken}}, \bibinfo {author} {\bibfnamefont {R.}~\bibnamefont {Barkhouser}}, \bibinfo {author} {\bibfnamefont {S.}~\bibnamefont {Bastian}}, \bibinfo {author} {\bibfnamefont {E.}~\bibnamefont {Berman}}, \emph {et~al.},\ }\bibfield  {title} {\bibinfo {title} {The sloan digital sky survey: Technical summary},\ }\href@noop {} {\bibfield  {journal} {\bibinfo  {journal} {The Astronomical Journal}\ }\textbf {\bibinfo {volume} {120}},\ \bibinfo {pages} {1579} (\bibinfo {year} {2000})}\BibitemShut {NoStop}%
\bibitem [{\citenamefont {Dewdney}\ \emph {et~al.}(2009)\citenamefont {Dewdney}, \citenamefont {Hall}, \citenamefont {Schilizzi},\ and\ \citenamefont {Lazio}}]{dewdney2009square}%
  \BibitemOpen
  \bibfield  {author} {\bibinfo {author} {\bibfnamefont {P.~E.}\ \bibnamefont {Dewdney}}, \bibinfo {author} {\bibfnamefont {P.~J.}\ \bibnamefont {Hall}}, \bibinfo {author} {\bibfnamefont {R.~T.}\ \bibnamefont {Schilizzi}},\ and\ \bibinfo {author} {\bibfnamefont {T.~J.~L.}\ \bibnamefont {Lazio}},\ }\bibfield  {title} {\bibinfo {title} {The square kilometre array},\ }\href@noop {} {\bibfield  {journal} {\bibinfo  {journal} {Proceedings of the IEEE}\ }\textbf {\bibinfo {volume} {97}},\ \bibinfo {pages} {1482} (\bibinfo {year} {2009})}\BibitemShut {NoStop}%
\bibitem [{\citenamefont {Razim}\ \emph {et~al.}(2021)\citenamefont {Razim}, \citenamefont {Cavuoti}, \citenamefont {Brescia}, \citenamefont {Riccio}, \citenamefont {Salvato},\ and\ \citenamefont {Longo}}]{razim2021improving}%
  \BibitemOpen
  \bibfield  {author} {\bibinfo {author} {\bibfnamefont {O.}~\bibnamefont {Razim}}, \bibinfo {author} {\bibfnamefont {S.}~\bibnamefont {Cavuoti}}, \bibinfo {author} {\bibfnamefont {M.}~\bibnamefont {Brescia}}, \bibinfo {author} {\bibfnamefont {G.}~\bibnamefont {Riccio}}, \bibinfo {author} {\bibfnamefont {M.}~\bibnamefont {Salvato}},\ and\ \bibinfo {author} {\bibfnamefont {G.}~\bibnamefont {Longo}},\ }\bibfield  {title} {\bibinfo {title} {Improving the reliability of photometric redshift with machine learning},\ }\href@noop {} {\bibfield  {journal} {\bibinfo  {journal} {Monthly Notices of the Royal Astronomical Society}\ }\textbf {\bibinfo {volume} {507}},\ \bibinfo {pages} {5034} (\bibinfo {year} {2021})}\BibitemShut {NoStop}%
\bibitem [{\citenamefont {Banerji}\ \emph {et~al.}(2010)\citenamefont {Banerji}, \citenamefont {Lahav}, \citenamefont {Lintott}, \citenamefont {Abdalla}, \citenamefont {Schawinski}, \citenamefont {Bamford}, \citenamefont {Andreescu}, \citenamefont {Murray}, \citenamefont {Raddick}, \citenamefont {Slosar} \emph {et~al.}}]{banerji2010galaxy}%
  \BibitemOpen
  \bibfield  {author} {\bibinfo {author} {\bibfnamefont {M.}~\bibnamefont {Banerji}}, \bibinfo {author} {\bibfnamefont {O.}~\bibnamefont {Lahav}}, \bibinfo {author} {\bibfnamefont {C.~J.}\ \bibnamefont {Lintott}}, \bibinfo {author} {\bibfnamefont {F.~B.}\ \bibnamefont {Abdalla}}, \bibinfo {author} {\bibfnamefont {K.}~\bibnamefont {Schawinski}}, \bibinfo {author} {\bibfnamefont {S.~P.}\ \bibnamefont {Bamford}}, \bibinfo {author} {\bibfnamefont {D.}~\bibnamefont {Andreescu}}, \bibinfo {author} {\bibfnamefont {P.}~\bibnamefont {Murray}}, \bibinfo {author} {\bibfnamefont {M.~J.}\ \bibnamefont {Raddick}}, \bibinfo {author} {\bibfnamefont {A.}~\bibnamefont {Slosar}}, \emph {et~al.},\ }\bibfield  {title} {\bibinfo {title} {Galaxy zoo: reproducing galaxy morphologies via machine learning},\ }\href@noop {} {\bibfield  {journal} {\bibinfo  {journal} {Monthly Notices of the Royal Astronomical Society}\ }\textbf {\bibinfo {volume} {406}},\ \bibinfo {pages} {342} (\bibinfo {year} {2010})}\BibitemShut {NoStop}%
\bibitem [{\citenamefont {Fluke}\ and\ \citenamefont {Jacobs}(2020)}]{fluke2020surveying}%
  \BibitemOpen
  \bibfield  {author} {\bibinfo {author} {\bibfnamefont {C.~J.}\ \bibnamefont {Fluke}}\ and\ \bibinfo {author} {\bibfnamefont {C.}~\bibnamefont {Jacobs}},\ }\bibfield  {title} {\bibinfo {title} {Surveying the reach and maturity of machine learning and artificial intelligence in astronomy},\ }\href@noop {} {\bibfield  {journal} {\bibinfo  {journal} {Wiley Interdisciplinary Reviews: Data Mining and Knowledge Discovery}\ }\textbf {\bibinfo {volume} {10}},\ \bibinfo {pages} {e1349} (\bibinfo {year} {2020})}\BibitemShut {NoStop}%
\bibitem [{\citenamefont {Baron}(2019)}]{baron2019machine}%
  \BibitemOpen
  \bibfield  {author} {\bibinfo {author} {\bibfnamefont {D.}~\bibnamefont {Baron}},\ }\bibfield  {title} {\bibinfo {title} {Machine learning in astronomy: A practical overview},\ }\href@noop {} {\bibfield  {journal} {\bibinfo  {journal} {arXiv preprint arXiv:1904.07248}\ } (\bibinfo {year} {2019})}\BibitemShut {NoStop}%
\bibitem [{\citenamefont {Slabbert}\ \emph {et~al.}(2024)\citenamefont {Slabbert}, \citenamefont {Lourens},\ and\ \citenamefont {Petruccione}}]{slabbert2024pulsar}%
  \BibitemOpen
  \bibfield  {author} {\bibinfo {author} {\bibfnamefont {D.}~\bibnamefont {Slabbert}}, \bibinfo {author} {\bibfnamefont {M.}~\bibnamefont {Lourens}},\ and\ \bibinfo {author} {\bibfnamefont {F.}~\bibnamefont {Petruccione}},\ }\bibfield  {title} {\bibinfo {title} {Pulsar classification: comparing quantum convolutional neural networks and quantum support vector machines},\ }\href@noop {} {\bibfield  {journal} {\bibinfo  {journal} {Quantum Machine Intelligence}\ }\textbf {\bibinfo {volume} {6}},\ \bibinfo {pages} {56} (\bibinfo {year} {2024})}\BibitemShut {NoStop}%
\bibitem [{\citenamefont {Keith}\ \emph {et~al.}(2010)\citenamefont {Keith}, \citenamefont {Jameson}, \citenamefont {Van~Straten}, \citenamefont {Bailes}, \citenamefont {Johnston}, \citenamefont {Kramer}, \citenamefont {Possenti}, \citenamefont {Bates}, \citenamefont {Bhat}, \citenamefont {Burgay} \emph {et~al.}}]{keith2010high}%
  \BibitemOpen
  \bibfield  {author} {\bibinfo {author} {\bibfnamefont {M.}~\bibnamefont {Keith}}, \bibinfo {author} {\bibfnamefont {A.}~\bibnamefont {Jameson}}, \bibinfo {author} {\bibfnamefont {W.}~\bibnamefont {Van~Straten}}, \bibinfo {author} {\bibfnamefont {M.}~\bibnamefont {Bailes}}, \bibinfo {author} {\bibfnamefont {S.}~\bibnamefont {Johnston}}, \bibinfo {author} {\bibfnamefont {M.}~\bibnamefont {Kramer}}, \bibinfo {author} {\bibfnamefont {A.}~\bibnamefont {Possenti}}, \bibinfo {author} {\bibfnamefont {S.}~\bibnamefont {Bates}}, \bibinfo {author} {\bibfnamefont {N.}~\bibnamefont {Bhat}}, \bibinfo {author} {\bibfnamefont {M.}~\bibnamefont {Burgay}}, \emph {et~al.},\ }\bibfield  {title} {\bibinfo {title} {The high time resolution universe pulsar survey--i. system configuration and initial discoveries},\ }\href@noop {} {\bibfield  {journal} {\bibinfo  {journal} {Monthly Notices of the Royal Astronomical Society}\ }\textbf {\bibinfo {volume} {409}},\ \bibinfo {pages} {619} (\bibinfo {year} {2010})}\BibitemShut {NoStop}%
\bibitem [{\citenamefont {Lyon}(2015)}]{htru2_372}%
  \BibitemOpen
  \bibfield  {author} {\bibinfo {author} {\bibfnamefont {R.}~\bibnamefont {Lyon}},\ }\href@noop {} {\bibinfo {title} {{HTRU2}}},\ \bibinfo {howpublished} {UCI Machine Learning Repository} (\bibinfo {year} {2015}),\ \bibinfo {note} {{DOI}: https://doi.org/10.24432/C5DK6R}\BibitemShut {NoStop}%
\bibitem [{\citenamefont {Morello}\ \emph {et~al.}(2014)\citenamefont {Morello}, \citenamefont {Barr}, \citenamefont {Bailes}, \citenamefont {Flynn}, \citenamefont {Keane},\ and\ \citenamefont {Van~Straten}}]{morello2014spinn}%
  \BibitemOpen
  \bibfield  {author} {\bibinfo {author} {\bibfnamefont {V.}~\bibnamefont {Morello}}, \bibinfo {author} {\bibfnamefont {E.}~\bibnamefont {Barr}}, \bibinfo {author} {\bibfnamefont {M.}~\bibnamefont {Bailes}}, \bibinfo {author} {\bibfnamefont {C.}~\bibnamefont {Flynn}}, \bibinfo {author} {\bibfnamefont {E.}~\bibnamefont {Keane}},\ and\ \bibinfo {author} {\bibfnamefont {W.}~\bibnamefont {Van~Straten}},\ }\bibfield  {title} {\bibinfo {title} {Spinn: a straightforward machine learning solution to the pulsar candidate selection problem},\ }\href@noop {} {\bibfield  {journal} {\bibinfo  {journal} {Monthly Notices of the Royal Astronomical Society}\ }\textbf {\bibinfo {volume} {443}},\ \bibinfo {pages} {1651} (\bibinfo {year} {2014})}\BibitemShut {NoStop}%
\bibitem [{\citenamefont {Lyne}\ and\ \citenamefont {Graham-Smith}(2012)}]{lyne2012pulsar}%
  \BibitemOpen
  \bibfield  {author} {\bibinfo {author} {\bibfnamefont {A.}~\bibnamefont {Lyne}}\ and\ \bibinfo {author} {\bibfnamefont {F.}~\bibnamefont {Graham-Smith}},\ }\href {https://doi.org/10.1017/CBO9780511844584} {\emph {\bibinfo {title} {Pulsar Astronomy}}},\ \bibinfo {edition} {4th}\ ed.,\ Cambridge Astrophysics\ (\bibinfo  {publisher} {Cambridge University Press},\ \bibinfo {year} {2012})\BibitemShut {NoStop}%
\bibitem [{\citenamefont {Kippenhahn}\ \emph {et~al.}(2012)\citenamefont {Kippenhahn}, \citenamefont {Weigert},\ and\ \citenamefont {Weiss}}]{kippenhahn2012stellar}%
  \BibitemOpen
  \bibfield  {author} {\bibinfo {author} {\bibfnamefont {R.}~\bibnamefont {Kippenhahn}}, \bibinfo {author} {\bibfnamefont {A.}~\bibnamefont {Weigert}},\ and\ \bibinfo {author} {\bibfnamefont {A.}~\bibnamefont {Weiss}},\ }\href {https://doi.org/10.1007/978-3-642-30304-3} {\emph {\bibinfo {title} {Stellar Structure and Evolution}}},\ \bibinfo {edition} {2nd}\ ed.,\ Astronomy and Astrophysics Library\ (\bibinfo  {publisher} {Springer Berlin, Heidelberg},\ \bibinfo {year} {2012})\BibitemShut {NoStop}%
\bibitem [{\citenamefont {Hulse}\ and\ \citenamefont {Taylor}(1975)}]{hulse1975discovery}%
  \BibitemOpen
  \bibfield  {author} {\bibinfo {author} {\bibfnamefont {R.~A.}\ \bibnamefont {Hulse}}\ and\ \bibinfo {author} {\bibfnamefont {J.~H.}\ \bibnamefont {Taylor}},\ }\bibfield  {title} {\bibinfo {title} {Discovery of a pulsar in a binary system},\ }\href {https://doi.org/10.1086/181708} {\bibfield  {journal} {\bibinfo  {journal} {The Astrophysical Journal}\ }\textbf {\bibinfo {volume} {195}},\ \bibinfo {pages} {L51} (\bibinfo {year} {1975})}\BibitemShut {NoStop}%
\bibitem [{\citenamefont {Stairs}(2003)}]{stairs2003testing}%
  \BibitemOpen
  \bibfield  {author} {\bibinfo {author} {\bibfnamefont {I.~H.}\ \bibnamefont {Stairs}},\ }\bibfield  {title} {\bibinfo {title} {Testing general relativity with pulsar timing},\ }\href {https://doi.org/10.12942/lrr-2003-5} {\bibfield  {journal} {\bibinfo  {journal} {Living Reviews in Relativity}\ }\textbf {\bibinfo {volume} {6}},\ \bibinfo {pages} {1} (\bibinfo {year} {2003})}\BibitemShut {NoStop}%
\bibitem [{\citenamefont {Foster}(1990)}]{foster1990constructing}%
  \BibitemOpen
  \bibfield  {author} {\bibinfo {author} {\bibfnamefont {I.}~\bibnamefont {Foster}, \bibfnamefont {Roger~Sherman}},\ }\emph {\bibinfo {title} {Constructing a pulsar timing array}},\ \href {http://ez.sun.ac.za/login?url=https://www.proquest.com/dissertations-theses/constructing-pulsar-timing-array/docview/303811082/se-2?accountid=14049} {Ph.D. thesis},\ \bibinfo  {school} {University of California, Berkeley}, \bibinfo {address} {Ann Arbor} (\bibinfo {year} {1990})\BibitemShut {NoStop}%
\bibitem [{\citenamefont {McLaughlin}(2013)}]{mclaughlin2013north}%
  \BibitemOpen
  \bibfield  {author} {\bibinfo {author} {\bibfnamefont {M.~A.}\ \bibnamefont {McLaughlin}},\ }\bibfield  {title} {\bibinfo {title} {The north american nanohertz observatory for gravitational waves},\ }\href {https://doi.org/10.1088/0264-9381/30/22/224008} {\bibfield  {journal} {\bibinfo  {journal} {Classical and Quantum Gravity}\ }\textbf {\bibinfo {volume} {30}},\ \bibinfo {pages} {224008} (\bibinfo {year} {2013})}\BibitemShut {NoStop}%
\bibitem [{\citenamefont {Verbiest}\ \emph {et~al.}(2016)\citenamefont {Verbiest}, \citenamefont {Lentati}, \citenamefont {Hobbs}, \citenamefont {van Haasteren}, \citenamefont {Demorest}, \citenamefont {Janssen}, \citenamefont {Wang}, \citenamefont {Desvignes}, \citenamefont {Caballero}, \citenamefont {Keith} \emph {et~al.}}]{verbiest2016international}%
  \BibitemOpen
  \bibfield  {author} {\bibinfo {author} {\bibfnamefont {J.}~\bibnamefont {Verbiest}}, \bibinfo {author} {\bibfnamefont {L.}~\bibnamefont {Lentati}}, \bibinfo {author} {\bibfnamefont {G.}~\bibnamefont {Hobbs}}, \bibinfo {author} {\bibfnamefont {R.}~\bibnamefont {van Haasteren}}, \bibinfo {author} {\bibfnamefont {P.~B.}\ \bibnamefont {Demorest}}, \bibinfo {author} {\bibfnamefont {G.}~\bibnamefont {Janssen}}, \bibinfo {author} {\bibfnamefont {J.-B.}\ \bibnamefont {Wang}}, \bibinfo {author} {\bibfnamefont {G.}~\bibnamefont {Desvignes}}, \bibinfo {author} {\bibfnamefont {R.}~\bibnamefont {Caballero}}, \bibinfo {author} {\bibfnamefont {M.}~\bibnamefont {Keith}}, \emph {et~al.},\ }\bibfield  {title} {\bibinfo {title} {The international pulsar timing array: first data release},\ }\href {https://doi.org/10.1093/mnras/stw347} {\bibfield  {journal} {\bibinfo  {journal} {Monthly Notices of the Royal Astronomical Society}\ }\textbf {\bibinfo {volume} {458}},\ \bibinfo {pages} {1267} (\bibinfo {year} {2016})}\BibitemShut
  {NoStop}%
\bibitem [{\citenamefont {Schuld}(2021{\natexlab{a}})}]{MariaSchuld2021}%
  \BibitemOpen
  \bibfield  {author} {\bibinfo {author} {\bibfnamefont {M.}~\bibnamefont {Schuld}},\ }\href@noop {} {\bibinfo {title} {Kernel-based training of quantum models with scikit-learn}},\ \bibinfo {howpublished} {\url{https://pennylane.ai/qml/demos/tutorial_kernel_based_training/}} (\bibinfo {year} {2021}{\natexlab{a}}),\ \bibinfo {note} {date Accessed: 2024-10-08}\BibitemShut {NoStop}%
\bibitem [{\citenamefont {Schuld}(2023)}]{schuld2023supervised}%
  \BibitemOpen
  \bibfield  {author} {\bibinfo {author} {\bibfnamefont {M.}~\bibnamefont {Schuld}},\ }\bibfield  {title} {\bibinfo {title} {Supervised quantum machine learning models are kernel methods (2021)},\ }\href@noop {} {\bibfield  {journal} {\bibinfo  {journal} {arXiv preprint arXiv:2101.11020}\ } (\bibinfo {year} {2023})}\BibitemShut {NoStop}%
\bibitem [{\citenamefont {Chandola}\ \emph {et~al.}(2009)\citenamefont {Chandola}, \citenamefont {Banerjee},\ and\ \citenamefont {Kumar}}]{chandola2009anomaly}%
  \BibitemOpen
  \bibfield  {author} {\bibinfo {author} {\bibfnamefont {V.}~\bibnamefont {Chandola}}, \bibinfo {author} {\bibfnamefont {A.}~\bibnamefont {Banerjee}},\ and\ \bibinfo {author} {\bibfnamefont {V.}~\bibnamefont {Kumar}},\ }\bibfield  {title} {\bibinfo {title} {Anomaly detection: A survey},\ }\href@noop {} {\bibfield  {journal} {\bibinfo  {journal} {ACM computing surveys (CSUR)}\ }\textbf {\bibinfo {volume} {41}},\ \bibinfo {pages} {1} (\bibinfo {year} {2009})}\BibitemShut {NoStop}%
\bibitem [{\citenamefont {LeCun}\ \emph {et~al.}(1998)\citenamefont {LeCun}, \citenamefont {Bottou}, \citenamefont {Bengio},\ and\ \citenamefont {Haffner}}]{lecun1998gradient}%
  \BibitemOpen
  \bibfield  {author} {\bibinfo {author} {\bibfnamefont {Y.}~\bibnamefont {LeCun}}, \bibinfo {author} {\bibfnamefont {L.}~\bibnamefont {Bottou}}, \bibinfo {author} {\bibfnamefont {Y.}~\bibnamefont {Bengio}},\ and\ \bibinfo {author} {\bibfnamefont {P.}~\bibnamefont {Haffner}},\ }\bibfield  {title} {\bibinfo {title} {Gradient-based learning applied to document recognition},\ }\href@noop {} {\bibfield  {journal} {\bibinfo  {journal} {Proceedings of the IEEE}\ }\textbf {\bibinfo {volume} {86}},\ \bibinfo {pages} {2278} (\bibinfo {year} {1998})}\BibitemShut {NoStop}%
\bibitem [{\citenamefont {Krizhevsky}\ and\ \citenamefont {Hinton}(2009)}]{krizhevsky2009learning}%
  \BibitemOpen
  \bibfield  {author} {\bibinfo {author} {\bibfnamefont {A.}~\bibnamefont {Krizhevsky}}\ and\ \bibinfo {author} {\bibfnamefont {G.}~\bibnamefont {Hinton}},\ }\href@noop {} {\emph {\bibinfo {title} {Learning multiple layers of features from tiny images}}},\ \bibinfo {type} {Technical Report}\ \bibinfo {number} {UTML TR 2009-003}\ (\bibinfo  {institution} {University of Toronto},\ \bibinfo {year} {2009})\BibitemShut {NoStop}%
\bibitem [{\citenamefont {Lyon}\ \emph {et~al.}(2016)\citenamefont {Lyon}, \citenamefont {Stappers}, \citenamefont {Cooper}, \citenamefont {Brooke},\ and\ \citenamefont {Knowles}}]{lyon2016fifty}%
  \BibitemOpen
  \bibfield  {author} {\bibinfo {author} {\bibfnamefont {R.~J.}\ \bibnamefont {Lyon}}, \bibinfo {author} {\bibfnamefont {B.}~\bibnamefont {Stappers}}, \bibinfo {author} {\bibfnamefont {S.}~\bibnamefont {Cooper}}, \bibinfo {author} {\bibfnamefont {J.~M.}\ \bibnamefont {Brooke}},\ and\ \bibinfo {author} {\bibfnamefont {J.~D.}\ \bibnamefont {Knowles}},\ }\bibfield  {title} {\bibinfo {title} {Fifty years of pulsar candidate selection: from simple filters to a new principled real-time classification approach},\ }\href {https://doi.org/10.1093/mnras/stw656} {\bibfield  {journal} {\bibinfo  {journal} {Monthly Notices of the Royal Astronomical Society}\ }\textbf {\bibinfo {volume} {459}},\ \bibinfo {pages} {1104} (\bibinfo {year} {2016})}\BibitemShut {NoStop}%
\bibitem [{\citenamefont {Pande}(2024)}]{pande2024comprehensive}%
  \BibitemOpen
  \bibfield  {author} {\bibinfo {author} {\bibfnamefont {M.~B.}\ \bibnamefont {Pande}},\ }\bibfield  {title} {\bibinfo {title} {A comprehensive review of data encoding techniques for quantum machine learning problems},\ }in\ \href@noop {} {\emph {\bibinfo {booktitle} {2024 Second International Conference on Emerging Trends in Information Technology and Engineering (ICETITE)}}}\ (\bibinfo {organization} {IEEE},\ \bibinfo {year} {2024})\ pp.\ \bibinfo {pages} {1--7}\BibitemShut {NoStop}%
\bibitem [{\citenamefont {Schuld}\ \emph {et~al.}(2021)\citenamefont {Schuld}, \citenamefont {Sweke},\ and\ \citenamefont {Meyer}}]{schuld2021effect}%
  \BibitemOpen
  \bibfield  {author} {\bibinfo {author} {\bibfnamefont {M.}~\bibnamefont {Schuld}}, \bibinfo {author} {\bibfnamefont {R.}~\bibnamefont {Sweke}},\ and\ \bibinfo {author} {\bibfnamefont {J.~J.}\ \bibnamefont {Meyer}},\ }\bibfield  {title} {\bibinfo {title} {Effect of data encoding on the expressive power of variational quantum-machine-learning models},\ }\href@noop {} {\bibfield  {journal} {\bibinfo  {journal} {Physical Review A}\ }\textbf {\bibinfo {volume} {103}},\ \bibinfo {pages} {032430} (\bibinfo {year} {2021})}\BibitemShut {NoStop}%
\bibitem [{\citenamefont {Ziou}\ and\ \citenamefont {Tabbone}(1998)}]{ziou1998edge}%
  \BibitemOpen
  \bibfield  {author} {\bibinfo {author} {\bibfnamefont {D.}~\bibnamefont {Ziou}}\ and\ \bibinfo {author} {\bibfnamefont {S.}~\bibnamefont {Tabbone}},\ }\bibfield  {title} {\bibinfo {title} {Edge detection techniques-an overview},\ }\href@noop {} {\bibfield  {journal} {\bibinfo  {journal} {Pattern Recognition and Image Analysis: Advances in Mathematical Theory and Applications}\ }\textbf {\bibinfo {volume} {8}},\ \bibinfo {pages} {537} (\bibinfo {year} {1998})}\BibitemShut {NoStop}%
\bibitem [{\citenamefont {Dalal}\ and\ \citenamefont {Triggs}(2005)}]{dalal2005histograms}%
  \BibitemOpen
  \bibfield  {author} {\bibinfo {author} {\bibfnamefont {N.}~\bibnamefont {Dalal}}\ and\ \bibinfo {author} {\bibfnamefont {B.}~\bibnamefont {Triggs}},\ }\bibfield  {title} {\bibinfo {title} {Histograms of oriented gradients for human detection},\ }in\ \href@noop {} {\emph {\bibinfo {booktitle} {2005 IEEE computer society conference on computer vision and pattern recognition (CVPR'05)}}},\ Vol.~\bibinfo {volume} {1}\ (\bibinfo {organization} {Ieee},\ \bibinfo {year} {2005})\ pp.\ \bibinfo {pages} {886--893}\BibitemShut {NoStop}%
\bibitem [{\citenamefont {Wold}\ \emph {et~al.}(1987)\citenamefont {Wold}, \citenamefont {Esbensen},\ and\ \citenamefont {Geladi}}]{wold1987principal}%
  \BibitemOpen
  \bibfield  {author} {\bibinfo {author} {\bibfnamefont {S.}~\bibnamefont {Wold}}, \bibinfo {author} {\bibfnamefont {K.}~\bibnamefont {Esbensen}},\ and\ \bibinfo {author} {\bibfnamefont {P.}~\bibnamefont {Geladi}},\ }\bibfield  {title} {\bibinfo {title} {Principal component analysis},\ }\href@noop {} {\bibfield  {journal} {\bibinfo  {journal} {Chemometrics and intelligent laboratory systems}\ }\textbf {\bibinfo {volume} {2}},\ \bibinfo {pages} {37} (\bibinfo {year} {1987})}\BibitemShut {NoStop}%
\bibitem [{\citenamefont {Pearson}(1901)}]{pearson1901liii}%
  \BibitemOpen
  \bibfield  {author} {\bibinfo {author} {\bibfnamefont {K.}~\bibnamefont {Pearson}},\ }\bibfield  {title} {\bibinfo {title} {{LIII. O}n lines and planes of closest fit to systems of points in space},\ }\href@noop {} {\bibfield  {journal} {\bibinfo  {journal} {The London, Edinburgh, and Dublin philosophical magazine and journal of science}\ }\textbf {\bibinfo {volume} {2}},\ \bibinfo {pages} {559} (\bibinfo {year} {1901})}\BibitemShut {NoStop}%
\bibitem [{\citenamefont {Jogin}\ \emph {et~al.}(2018)\citenamefont {Jogin}, \citenamefont {Madhulika}, \citenamefont {Divya}, \citenamefont {Meghana}, \citenamefont {Apoorva} \emph {et~al.}}]{jogin2018feature}%
  \BibitemOpen
  \bibfield  {author} {\bibinfo {author} {\bibfnamefont {M.}~\bibnamefont {Jogin}}, \bibinfo {author} {\bibfnamefont {M.}~\bibnamefont {Madhulika}}, \bibinfo {author} {\bibfnamefont {G.}~\bibnamefont {Divya}}, \bibinfo {author} {\bibfnamefont {R.}~\bibnamefont {Meghana}}, \bibinfo {author} {\bibfnamefont {S.}~\bibnamefont {Apoorva}}, \emph {et~al.},\ }\bibfield  {title} {\bibinfo {title} {Feature extraction using convolution neural networks (cnn) and deep learning},\ }in\ \href@noop {} {\emph {\bibinfo {booktitle} {2018 3rd IEEE international conference on recent trends in electronics, information \& communication technology (RTEICT)}}}\ (\bibinfo {organization} {IEEE},\ \bibinfo {year} {2018})\ pp.\ \bibinfo {pages} {2319--2323}\BibitemShut {NoStop}%
\bibitem [{\citenamefont {Hinton}\ and\ \citenamefont {Salakhutdinov}(2006)}]{hinton2006reducing}%
  \BibitemOpen
  \bibfield  {author} {\bibinfo {author} {\bibfnamefont {G.~E.}\ \bibnamefont {Hinton}}\ and\ \bibinfo {author} {\bibfnamefont {R.~R.}\ \bibnamefont {Salakhutdinov}},\ }\bibfield  {title} {\bibinfo {title} {Reducing the dimensionality of data with neural networks},\ }\href@noop {} {\bibfield  {journal} {\bibinfo  {journal} {science}\ }\textbf {\bibinfo {volume} {313}},\ \bibinfo {pages} {504} (\bibinfo {year} {2006})}\BibitemShut {NoStop}%
\bibitem [{\citenamefont {Bl{\'a}zquez-Garc{\'\i}a}\ \emph {et~al.}(2021)\citenamefont {Bl{\'a}zquez-Garc{\'\i}a}, \citenamefont {Conde}, \citenamefont {Mori},\ and\ \citenamefont {Lozano}}]{blazquez2021review}%
  \BibitemOpen
  \bibfield  {author} {\bibinfo {author} {\bibfnamefont {A.}~\bibnamefont {Bl{\'a}zquez-Garc{\'\i}a}}, \bibinfo {author} {\bibfnamefont {A.}~\bibnamefont {Conde}}, \bibinfo {author} {\bibfnamefont {U.}~\bibnamefont {Mori}},\ and\ \bibinfo {author} {\bibfnamefont {J.~A.}\ \bibnamefont {Lozano}},\ }\bibfield  {title} {\bibinfo {title} {A review on outlier/anomaly detection in time series data},\ }\href@noop {} {\bibfield  {journal} {\bibinfo  {journal} {ACM computing surveys (CSUR)}\ }\textbf {\bibinfo {volume} {54}},\ \bibinfo {pages} {1} (\bibinfo {year} {2021})}\BibitemShut {NoStop}%
\bibitem [{\citenamefont {Hodge}\ and\ \citenamefont {Austin}(2004)}]{hodge2004survey}%
  \BibitemOpen
  \bibfield  {author} {\bibinfo {author} {\bibfnamefont {V.}~\bibnamefont {Hodge}}\ and\ \bibinfo {author} {\bibfnamefont {J.}~\bibnamefont {Austin}},\ }\bibfield  {title} {\bibinfo {title} {A survey of outlier detection methodologies},\ }\href@noop {} {\bibfield  {journal} {\bibinfo  {journal} {Artificial intelligence review}\ }\textbf {\bibinfo {volume} {22}},\ \bibinfo {pages} {85} (\bibinfo {year} {2004})}\BibitemShut {NoStop}%
\bibitem [{\citenamefont {He}\ and\ \citenamefont {Garcia}(2009)}]{he2009learning}%
  \BibitemOpen
  \bibfield  {author} {\bibinfo {author} {\bibfnamefont {H.}~\bibnamefont {He}}\ and\ \bibinfo {author} {\bibfnamefont {E.~A.}\ \bibnamefont {Garcia}},\ }\bibfield  {title} {\bibinfo {title} {Learning from imbalanced data},\ }\href@noop {} {\bibfield  {journal} {\bibinfo  {journal} {IEEE Transactions on knowledge and data engineering}\ }\textbf {\bibinfo {volume} {21}},\ \bibinfo {pages} {1263} (\bibinfo {year} {2009})}\BibitemShut {NoStop}%
\bibitem [{\citenamefont {Lima}\ \emph {et~al.}(2010)\citenamefont {Lima}, \citenamefont {Zarpelao}, \citenamefont {Sampaio}, \citenamefont {Rodrigues}, \citenamefont {Abrao},\ and\ \citenamefont {Proen{\c{c}}a}}]{lima2010anomaly}%
  \BibitemOpen
  \bibfield  {author} {\bibinfo {author} {\bibfnamefont {M.~F.}\ \bibnamefont {Lima}}, \bibinfo {author} {\bibfnamefont {B.~B.}\ \bibnamefont {Zarpelao}}, \bibinfo {author} {\bibfnamefont {L.~D.}\ \bibnamefont {Sampaio}}, \bibinfo {author} {\bibfnamefont {J.~J.}\ \bibnamefont {Rodrigues}}, \bibinfo {author} {\bibfnamefont {T.}~\bibnamefont {Abrao}},\ and\ \bibinfo {author} {\bibfnamefont {M.~L.}\ \bibnamefont {Proen{\c{c}}a}},\ }\bibfield  {title} {\bibinfo {title} {Anomaly detection using baseline and k-means clustering},\ }in\ \href@noop {} {\emph {\bibinfo {booktitle} {SoftCOM 2010, 18th International Conference on Software, Telecommunications and Computer Networks}}}\ (\bibinfo {organization} {IEEE},\ \bibinfo {year} {2010})\ pp.\ \bibinfo {pages} {305--309}\BibitemShut {NoStop}%
\bibitem [{\citenamefont {Nachman}\ and\ \citenamefont {Shih}(2020)}]{nachman2020anomaly}%
  \BibitemOpen
  \bibfield  {author} {\bibinfo {author} {\bibfnamefont {B.}~\bibnamefont {Nachman}}\ and\ \bibinfo {author} {\bibfnamefont {D.}~\bibnamefont {Shih}},\ }\bibfield  {title} {\bibinfo {title} {Anomaly detection with density estimation},\ }\href@noop {} {\bibfield  {journal} {\bibinfo  {journal} {Physical Review D}\ }\textbf {\bibinfo {volume} {101}},\ \bibinfo {pages} {075042} (\bibinfo {year} {2020})}\BibitemShut {NoStop}%
\bibitem [{\citenamefont {Aissa}\ and\ \citenamefont {Guerroumi}(2016)}]{aissa2016semi}%
  \BibitemOpen
  \bibfield  {author} {\bibinfo {author} {\bibfnamefont {N.~B.}\ \bibnamefont {Aissa}}\ and\ \bibinfo {author} {\bibfnamefont {M.}~\bibnamefont {Guerroumi}},\ }\bibfield  {title} {\bibinfo {title} {Semi-supervised statistical approach for network anomaly detection},\ }\href@noop {} {\bibfield  {journal} {\bibinfo  {journal} {Procedia Computer Science}\ }\textbf {\bibinfo {volume} {83}},\ \bibinfo {pages} {1090} (\bibinfo {year} {2016})}\BibitemShut {NoStop}%
\bibitem [{\citenamefont {Liu}\ \emph {et~al.}(2008)\citenamefont {Liu}, \citenamefont {Ting},\ and\ \citenamefont {Zhou}}]{liu2008isolation}%
  \BibitemOpen
  \bibfield  {author} {\bibinfo {author} {\bibfnamefont {F.~T.}\ \bibnamefont {Liu}}, \bibinfo {author} {\bibfnamefont {K.~M.}\ \bibnamefont {Ting}},\ and\ \bibinfo {author} {\bibfnamefont {Z.-H.}\ \bibnamefont {Zhou}},\ }\bibfield  {title} {\bibinfo {title} {Isolation forest},\ }in\ \href@noop {} {\emph {\bibinfo {booktitle} {2008 eighth ieee international conference on data mining}}}\ (\bibinfo {organization} {IEEE},\ \bibinfo {year} {2008})\ pp.\ \bibinfo {pages} {413--422}\BibitemShut {NoStop}%
\bibitem [{\citenamefont {Powers}(2020)}]{powers2020evaluation}%
  \BibitemOpen
  \bibfield  {author} {\bibinfo {author} {\bibfnamefont {D.~M.}\ \bibnamefont {Powers}},\ }\bibfield  {title} {\bibinfo {title} {Evaluation: from precision, recall and f-measure to roc, informedness, markedness and correlation},\ }\bibfield  {journal} {\bibinfo  {journal} {arXiv preprint arXiv:2010.16061}\ }\href {https://doi.org/10.48550/arXiv.2010.16061} {10.48550/arXiv.2010.16061} (\bibinfo {year} {2020})\BibitemShut {NoStop}%
\bibitem [{\citenamefont {Tharwat}(2020)}]{tharwat2020classification}%
  \BibitemOpen
  \bibfield  {author} {\bibinfo {author} {\bibfnamefont {A.}~\bibnamefont {Tharwat}},\ }\bibfield  {title} {\bibinfo {title} {Classification assessment methods},\ }\href {https://doi.org/10.1016/j.aci.2018.08.003} {\bibfield  {journal} {\bibinfo  {journal} {Applied computing and informatics}\ }\textbf {\bibinfo {volume} {17}},\ \bibinfo {pages} {168} (\bibinfo {year} {2020})}\BibitemShut {NoStop}%
\bibitem [{\citenamefont {Chicco}\ and\ \citenamefont {Jurman}(2020)}]{chicco2020advantages}%
  \BibitemOpen
  \bibfield  {author} {\bibinfo {author} {\bibfnamefont {D.}~\bibnamefont {Chicco}}\ and\ \bibinfo {author} {\bibfnamefont {G.}~\bibnamefont {Jurman}},\ }\bibfield  {title} {\bibinfo {title} {The advantages of the matthews correlation coefficient (mcc) over f1 score and accuracy in binary classification evaluation},\ }\href@noop {} {\bibfield  {journal} {\bibinfo  {journal} {BMC genomics}\ }\textbf {\bibinfo {volume} {21}},\ \bibinfo {pages} {1} (\bibinfo {year} {2020})}\BibitemShut {NoStop}%
\bibitem [{\citenamefont {Noble}(2006)}]{noble2006support}%
  \BibitemOpen
  \bibfield  {author} {\bibinfo {author} {\bibfnamefont {W.~S.}\ \bibnamefont {Noble}},\ }\bibfield  {title} {\bibinfo {title} {What is a support vector machine?},\ }\href@noop {} {\bibfield  {journal} {\bibinfo  {journal} {Nature biotechnology}\ }\textbf {\bibinfo {volume} {24}},\ \bibinfo {pages} {1565} (\bibinfo {year} {2006})}\BibitemShut {NoStop}%
\bibitem [{\citenamefont {Boswell}(2002)}]{boswell2002introduction}%
  \BibitemOpen
  \bibfield  {author} {\bibinfo {author} {\bibfnamefont {D.}~\bibnamefont {Boswell}},\ }\bibfield  {title} {\bibinfo {title} {Introduction to support vector machines},\ }\href@noop {} {\bibfield  {journal} {\bibinfo  {journal} {Departement of Computer Science and Engineering University of California San Diego}\ }\textbf {\bibinfo {volume} {11}},\ \bibinfo {pages} {16} (\bibinfo {year} {2002})}\BibitemShut {NoStop}%
\bibitem [{\citenamefont {Sch{\"o}lkopf}\ \emph {et~al.}(1999)\citenamefont {Sch{\"o}lkopf}, \citenamefont {Burges},\ and\ \citenamefont {Smola}}]{scholkopf1999advances}%
  \BibitemOpen
  \bibfield  {author} {\bibinfo {author} {\bibfnamefont {B.}~\bibnamefont {Sch{\"o}lkopf}}, \bibinfo {author} {\bibfnamefont {C.}~\bibnamefont {Burges}},\ and\ \bibinfo {author} {\bibfnamefont {A.}~\bibnamefont {Smola}},\ }\href {https://dl.acm.org/doi/10.5555/299094} {\emph {\bibinfo {title} {Advances in Kernel Methods: Support Vector Learning}}},\ Mit Press\ (\bibinfo  {publisher} {MIT Press},\ \bibinfo {year} {1999})\BibitemShut {NoStop}%
\bibitem [{\citenamefont {Boser}\ \emph {et~al.}(1992)\citenamefont {Boser}, \citenamefont {Guyon},\ and\ \citenamefont {Vapnik}}]{boser1992training}%
  \BibitemOpen
  \bibfield  {author} {\bibinfo {author} {\bibfnamefont {B.~E.}\ \bibnamefont {Boser}}, \bibinfo {author} {\bibfnamefont {I.~M.}\ \bibnamefont {Guyon}},\ and\ \bibinfo {author} {\bibfnamefont {V.~N.}\ \bibnamefont {Vapnik}},\ }\bibfield  {title} {\bibinfo {title} {A training algorithm for optimal margin classifiers},\ }in\ \href {https://doi.org/10.1145/130385.130401} {\emph {\bibinfo {booktitle} {Proceedings of the fifth annual workshop on Computational learning theory}}}\ (\bibinfo {year} {1992})\ pp.\ \bibinfo {pages} {144--152}\BibitemShut {NoStop}%
\bibitem [{\citenamefont {Sch{\"o}lkopf}\ and\ \citenamefont {Smola}(2002)}]{scholkopf2002learning}%
  \BibitemOpen
  \bibfield  {author} {\bibinfo {author} {\bibfnamefont {B.}~\bibnamefont {Sch{\"o}lkopf}}\ and\ \bibinfo {author} {\bibfnamefont {A.~J.}\ \bibnamefont {Smola}},\ }\href@noop {} {\emph {\bibinfo {title} {Learning with kernels: support vector machines, regularization, optimization, and beyond}}}\ (\bibinfo  {publisher} {MIT press},\ \bibinfo {year} {2002})\BibitemShut {NoStop}%
\bibitem [{\citenamefont {Sch{\"o}lkopf}\ \emph {et~al.}(2001)\citenamefont {Sch{\"o}lkopf}, \citenamefont {Platt}, \citenamefont {Shawe-Taylor}, \citenamefont {Smola},\ and\ \citenamefont {Williamson}}]{scholkopf2001estimating}%
  \BibitemOpen
  \bibfield  {author} {\bibinfo {author} {\bibfnamefont {B.}~\bibnamefont {Sch{\"o}lkopf}}, \bibinfo {author} {\bibfnamefont {J.~C.}\ \bibnamefont {Platt}}, \bibinfo {author} {\bibfnamefont {J.}~\bibnamefont {Shawe-Taylor}}, \bibinfo {author} {\bibfnamefont {A.~J.}\ \bibnamefont {Smola}},\ and\ \bibinfo {author} {\bibfnamefont {R.~C.}\ \bibnamefont {Williamson}},\ }\bibfield  {title} {\bibinfo {title} {Estimating the support of a high-dimensional distribution},\ }\href@noop {} {\bibfield  {journal} {\bibinfo  {journal} {Neural computation}\ }\textbf {\bibinfo {volume} {13}},\ \bibinfo {pages} {1443} (\bibinfo {year} {2001})}\BibitemShut {NoStop}%
\bibitem [{\citenamefont {M{\"u}ller}\ \emph {et~al.}(2018)\citenamefont {M{\"u}ller}, \citenamefont {Mika}, \citenamefont {Tsuda},\ and\ \citenamefont {Sch{\"o}lkopf}}]{muller2018introduction}%
  \BibitemOpen
  \bibfield  {author} {\bibinfo {author} {\bibfnamefont {K.-R.}\ \bibnamefont {M{\"u}ller}}, \bibinfo {author} {\bibfnamefont {S.}~\bibnamefont {Mika}}, \bibinfo {author} {\bibfnamefont {K.}~\bibnamefont {Tsuda}},\ and\ \bibinfo {author} {\bibfnamefont {K.}~\bibnamefont {Sch{\"o}lkopf}},\ }\bibfield  {title} {\bibinfo {title} {An introduction to kernel-based learning algorithms},\ }in\ \href@noop {} {\emph {\bibinfo {booktitle} {Handbook of neural network signal processing}}}\ (\bibinfo  {publisher} {CRC Press},\ \bibinfo {year} {2018})\ pp.\ \bibinfo {pages} {4--1}\BibitemShut {NoStop}%
\bibitem [{\citenamefont {Hejazi}\ and\ \citenamefont {Singh}(2013)}]{hejazi2013one}%
  \BibitemOpen
  \bibfield  {author} {\bibinfo {author} {\bibfnamefont {M.}~\bibnamefont {Hejazi}}\ and\ \bibinfo {author} {\bibfnamefont {Y.~P.}\ \bibnamefont {Singh}},\ }\bibfield  {title} {\bibinfo {title} {One-class support vector machines approach to anomaly detection},\ }\href@noop {} {\bibfield  {journal} {\bibinfo  {journal} {Applied Artificial Intelligence}\ }\textbf {\bibinfo {volume} {27}},\ \bibinfo {pages} {351} (\bibinfo {year} {2013})}\BibitemShut {NoStop}%
\bibitem [{\citenamefont {Huang}\ \emph {et~al.}(2021)\citenamefont {Huang}, \citenamefont {Broughton}, \citenamefont {Mohseni}, \citenamefont {Babbush}, \citenamefont {Boixo}, \citenamefont {Neven},\ and\ \citenamefont {McClean}}]{huang2021power}%
  \BibitemOpen
  \bibfield  {author} {\bibinfo {author} {\bibfnamefont {H.-Y.}\ \bibnamefont {Huang}}, \bibinfo {author} {\bibfnamefont {M.}~\bibnamefont {Broughton}}, \bibinfo {author} {\bibfnamefont {M.}~\bibnamefont {Mohseni}}, \bibinfo {author} {\bibfnamefont {R.}~\bibnamefont {Babbush}}, \bibinfo {author} {\bibfnamefont {S.}~\bibnamefont {Boixo}}, \bibinfo {author} {\bibfnamefont {H.}~\bibnamefont {Neven}},\ and\ \bibinfo {author} {\bibfnamefont {J.~R.}\ \bibnamefont {McClean}},\ }\bibfield  {title} {\bibinfo {title} {Power of data in quantum machine learning},\ }\href@noop {} {\bibfield  {journal} {\bibinfo  {journal} {Nature communications}\ }\textbf {\bibinfo {volume} {12}},\ \bibinfo {pages} {2631} (\bibinfo {year} {2021})}\BibitemShut {NoStop}%
\bibitem [{\citenamefont {Liu}\ \emph {et~al.}(2021)\citenamefont {Liu}, \citenamefont {Arunachalam},\ and\ \citenamefont {Temme}}]{liu2021rigorous}%
  \BibitemOpen
  \bibfield  {author} {\bibinfo {author} {\bibfnamefont {Y.}~\bibnamefont {Liu}}, \bibinfo {author} {\bibfnamefont {S.}~\bibnamefont {Arunachalam}},\ and\ \bibinfo {author} {\bibfnamefont {K.}~\bibnamefont {Temme}},\ }\bibfield  {title} {\bibinfo {title} {A rigorous and robust quantum speed-up in supervised machine learning},\ }\href@noop {} {\bibfield  {journal} {\bibinfo  {journal} {Nature Physics}\ }\textbf {\bibinfo {volume} {17}},\ \bibinfo {pages} {1013} (\bibinfo {year} {2021})}\BibitemShut {NoStop}%
\bibitem [{\citenamefont {Schuld}\ and\ \citenamefont {Killoran}(2019)}]{schuld2019quantum}%
  \BibitemOpen
  \bibfield  {author} {\bibinfo {author} {\bibfnamefont {M.}~\bibnamefont {Schuld}}\ and\ \bibinfo {author} {\bibfnamefont {N.}~\bibnamefont {Killoran}},\ }\bibfield  {title} {\bibinfo {title} {Quantum machine learning in feature hilbert spaces},\ }\href@noop {} {\bibfield  {journal} {\bibinfo  {journal} {Physical review letters}\ }\textbf {\bibinfo {volume} {122}},\ \bibinfo {pages} {040504} (\bibinfo {year} {2019})}\BibitemShut {NoStop}%
\bibitem [{\citenamefont {Havl{\'\i}{\v{c}}ek}\ \emph {et~al.}(2019)\citenamefont {Havl{\'\i}{\v{c}}ek}, \citenamefont {C{\'o}rcoles}, \citenamefont {Temme}, \citenamefont {Harrow}, \citenamefont {Kandala}, \citenamefont {Chow},\ and\ \citenamefont {Gambetta}}]{havlivcek2019supervised}%
  \BibitemOpen
  \bibfield  {author} {\bibinfo {author} {\bibfnamefont {V.}~\bibnamefont {Havl{\'\i}{\v{c}}ek}}, \bibinfo {author} {\bibfnamefont {A.~D.}\ \bibnamefont {C{\'o}rcoles}}, \bibinfo {author} {\bibfnamefont {K.}~\bibnamefont {Temme}}, \bibinfo {author} {\bibfnamefont {A.~W.}\ \bibnamefont {Harrow}}, \bibinfo {author} {\bibfnamefont {A.}~\bibnamefont {Kandala}}, \bibinfo {author} {\bibfnamefont {J.~M.}\ \bibnamefont {Chow}},\ and\ \bibinfo {author} {\bibfnamefont {J.~M.}\ \bibnamefont {Gambetta}},\ }\bibfield  {title} {\bibinfo {title} {Supervised learning with quantum-enhanced feature spaces},\ }\href@noop {} {\bibfield  {journal} {\bibinfo  {journal} {Nature}\ }\textbf {\bibinfo {volume} {567}},\ \bibinfo {pages} {209} (\bibinfo {year} {2019})}\BibitemShut {NoStop}%
\bibitem [{\citenamefont {Schuld}(2021{\natexlab{b}})}]{schuld2021supervised}%
  \BibitemOpen
  \bibfield  {author} {\bibinfo {author} {\bibfnamefont {M.}~\bibnamefont {Schuld}},\ }\bibfield  {title} {\bibinfo {title} {Supervised quantum machine learning models are kernel methods},\ }\href@noop {} {\bibfield  {journal} {\bibinfo  {journal} {arXiv preprint arXiv:2101.11020}\ } (\bibinfo {year} {2021}{\natexlab{b}})}\BibitemShut {NoStop}%
\bibitem [{\citenamefont {Goodfellow}(2016)}]{goodfellow2016deep}%
  \BibitemOpen
  \bibfield  {author} {\bibinfo {author} {\bibfnamefont {I.}~\bibnamefont {Goodfellow}},\ }\href@noop {} {\emph {\bibinfo {title} {Deep learning}}}\ (\bibinfo  {publisher} {MIT Press},\ \bibinfo {year} {2016})\BibitemShut {NoStop}%
\bibitem [{\citenamefont {Zhang}\ \emph {et~al.}(2021)\citenamefont {Zhang}, \citenamefont {Bengio}, \citenamefont {Hardt}, \citenamefont {Recht},\ and\ \citenamefont {Vinyals}}]{zhang2021understanding}%
  \BibitemOpen
  \bibfield  {author} {\bibinfo {author} {\bibfnamefont {C.}~\bibnamefont {Zhang}}, \bibinfo {author} {\bibfnamefont {S.}~\bibnamefont {Bengio}}, \bibinfo {author} {\bibfnamefont {M.}~\bibnamefont {Hardt}}, \bibinfo {author} {\bibfnamefont {B.}~\bibnamefont {Recht}},\ and\ \bibinfo {author} {\bibfnamefont {O.}~\bibnamefont {Vinyals}},\ }\bibfield  {title} {\bibinfo {title} {Understanding deep learning (still) requires rethinking generalization},\ }\href@noop {} {\bibfield  {journal} {\bibinfo  {journal} {Communications of the ACM}\ }\textbf {\bibinfo {volume} {64}},\ \bibinfo {pages} {107} (\bibinfo {year} {2021})}\BibitemShut {NoStop}%
\bibitem [{\citenamefont {Hochreiter}(1997)}]{hochreiter1997long}%
  \BibitemOpen
  \bibfield  {author} {\bibinfo {author} {\bibfnamefont {S.}~\bibnamefont {Hochreiter}},\ }\bibfield  {title} {\bibinfo {title} {Long short-term memory},\ }\href@noop {} {\bibfield  {journal} {\bibinfo  {journal} {Neural Computation MIT-Press}\ } (\bibinfo {year} {1997})}\BibitemShut {NoStop}%
\bibitem [{\citenamefont {Wang}\ \emph {et~al.}(2023)\citenamefont {Wang}, \citenamefont {Huang}, \citenamefont {Liu}, \citenamefont {Yi}, \citenamefont {Wu},\ and\ \citenamefont {Wang}}]{wang2023quantum}%
  \BibitemOpen
  \bibfield  {author} {\bibinfo {author} {\bibfnamefont {M.}~\bibnamefont {Wang}}, \bibinfo {author} {\bibfnamefont {A.}~\bibnamefont {Huang}}, \bibinfo {author} {\bibfnamefont {Y.}~\bibnamefont {Liu}}, \bibinfo {author} {\bibfnamefont {X.}~\bibnamefont {Yi}}, \bibinfo {author} {\bibfnamefont {J.}~\bibnamefont {Wu}},\ and\ \bibinfo {author} {\bibfnamefont {S.}~\bibnamefont {Wang}},\ }\bibfield  {title} {\bibinfo {title} {A quantum-classical hybrid solution for deep anomaly detection},\ }\href@noop {} {\bibfield  {journal} {\bibinfo  {journal} {Entropy}\ }\textbf {\bibinfo {volume} {25}},\ \bibinfo {pages} {427} (\bibinfo {year} {2023})}\BibitemShut {NoStop}%
\bibitem [{\citenamefont {Senokosov}\ \emph {et~al.}(2024)\citenamefont {Senokosov}, \citenamefont {Sedykh}, \citenamefont {Sagingalieva}, \citenamefont {Kyriacou},\ and\ \citenamefont {Melnikov}}]{senokosov2024quantum}%
  \BibitemOpen
  \bibfield  {author} {\bibinfo {author} {\bibfnamefont {A.}~\bibnamefont {Senokosov}}, \bibinfo {author} {\bibfnamefont {A.}~\bibnamefont {Sedykh}}, \bibinfo {author} {\bibfnamefont {A.}~\bibnamefont {Sagingalieva}}, \bibinfo {author} {\bibfnamefont {B.}~\bibnamefont {Kyriacou}},\ and\ \bibinfo {author} {\bibfnamefont {A.}~\bibnamefont {Melnikov}},\ }\bibfield  {title} {\bibinfo {title} {Quantum machine learning for image classification},\ }\href@noop {} {\bibfield  {journal} {\bibinfo  {journal} {Machine Learning: Science and Technology}\ }\textbf {\bibinfo {volume} {5}},\ \bibinfo {pages} {015040} (\bibinfo {year} {2024})}\BibitemShut {NoStop}%
\bibitem [{\citenamefont {He}\ \emph {et~al.}(2016)\citenamefont {He}, \citenamefont {Zhang}, \citenamefont {Ren},\ and\ \citenamefont {Sun}}]{he2016deep}%
  \BibitemOpen
  \bibfield  {author} {\bibinfo {author} {\bibfnamefont {K.}~\bibnamefont {He}}, \bibinfo {author} {\bibfnamefont {X.}~\bibnamefont {Zhang}}, \bibinfo {author} {\bibfnamefont {S.}~\bibnamefont {Ren}},\ and\ \bibinfo {author} {\bibfnamefont {J.}~\bibnamefont {Sun}},\ }\bibfield  {title} {\bibinfo {title} {Deep residual learning for image recognition},\ }in\ \href@noop {} {\emph {\bibinfo {booktitle} {Proceedings of the IEEE conference on computer vision and pattern recognition}}}\ (\bibinfo {year} {2016})\ pp.\ \bibinfo {pages} {770--778}\BibitemShut {NoStop}%
\bibitem [{\citenamefont {Gong}\ \emph {et~al.}(2022)\citenamefont {Gong}, \citenamefont {Liu}, \citenamefont {Pei}, \citenamefont {Wu},\ and\ \citenamefont {Guo}}]{gong2022resnet10}%
  \BibitemOpen
  \bibfield  {author} {\bibinfo {author} {\bibfnamefont {J.}~\bibnamefont {Gong}}, \bibinfo {author} {\bibfnamefont {W.}~\bibnamefont {Liu}}, \bibinfo {author} {\bibfnamefont {M.}~\bibnamefont {Pei}}, \bibinfo {author} {\bibfnamefont {C.}~\bibnamefont {Wu}},\ and\ \bibinfo {author} {\bibfnamefont {L.}~\bibnamefont {Guo}},\ }\bibfield  {title} {\bibinfo {title} {Resnet10: A lightweight residual network for remote sensing image classification},\ }in\ \href@noop {} {\emph {\bibinfo {booktitle} {2022 14th International Conference on Measuring Technology and Mechatronics Automation (ICMTMA)}}}\ (\bibinfo {organization} {IEEE},\ \bibinfo {year} {2022})\ pp.\ \bibinfo {pages} {975--978}\BibitemShut {NoStop}%
\bibitem [{\citenamefont {Wang}\ \emph {et~al.}(2004)\citenamefont {Wang}, \citenamefont {Bovik}, \citenamefont {Sheikh},\ and\ \citenamefont {Simoncelli}}]{wang2004image}%
  \BibitemOpen
  \bibfield  {author} {\bibinfo {author} {\bibfnamefont {Z.}~\bibnamefont {Wang}}, \bibinfo {author} {\bibfnamefont {A.~C.}\ \bibnamefont {Bovik}}, \bibinfo {author} {\bibfnamefont {H.~R.}\ \bibnamefont {Sheikh}},\ and\ \bibinfo {author} {\bibfnamefont {E.~P.}\ \bibnamefont {Simoncelli}},\ }\bibfield  {title} {\bibinfo {title} {Image quality assessment: from error visibility to structural similarity},\ }\href@noop {} {\bibfield  {journal} {\bibinfo  {journal} {IEEE transactions on image processing}\ }\textbf {\bibinfo {volume} {13}},\ \bibinfo {pages} {600} (\bibinfo {year} {2004})}\BibitemShut {NoStop}%
\bibitem [{\citenamefont {Dietterich}(2000)}]{dietterich2000ensemble}%
  \BibitemOpen
  \bibfield  {author} {\bibinfo {author} {\bibfnamefont {T.~G.}\ \bibnamefont {Dietterich}},\ }\bibfield  {title} {\bibinfo {title} {Ensemble methods in machine learning},\ }in\ \href@noop {} {\emph {\bibinfo {booktitle} {International workshop on multiple classifier systems}}}\ (\bibinfo {organization} {Springer},\ \bibinfo {year} {2000})\ pp.\ \bibinfo {pages} {1--15}\BibitemShut {NoStop}%
\end{thebibliography}%

\end{document}